\documentclass[11pt,a4paper]{article}
\usepackage{jcappub}

\usepackage{mathrsfs,amsmath,amssymb,graphicx}

\newcommand{\de}{{\mathrm d}}
\newcommand{\om}{{\Omega_m}}
\newcommand{\ob}{{\Omega_b}}
\newcommand{\odm}{{\Omega_\mathrm{DM}}}

\newcommand{\ol}{{\Omega_\Lambda}}
\newcommand{\ho}{{H_0}}
\newcommand{\lcdm}{$\Lambda$CDM}
\newcommand{\cinf}{{c_\infty}}

\title{Inclusive Constraints on Unified Dark Matter Models from Future Large-Scale Surveys}

\author[a]{Stefano Camera,}
\author[b,c,d]{Carmelita Carbone}
\author[b,c,d]{ and Lauro Moscardini}
\affiliation[a]{CENTRA, Instituto Superior T\'ecnico, Universidade T\'ecnica de Lisboa,\\Av. Rovisco Pais 1, 1049-001 Lisboa, Portugal}
\affiliation[b]{Dipartimento di Astronomia, Universit\`a di Bologna,\\V. Ranzani 1, 40127 Bologna, Italy}
\affiliation[c]{INAF, Osservatorio Astronomico di Bologna,\\V. Ranzani 1, 40127 Bologna, Italy}
\affiliation[d]{INFN, Sezione di Bologna,\\V. Ranzani 1, 40127 Bologna, Italy}
\emailAdd{stefano.camera@ist.utl.pt}
\emailAdd{carmelita.carbone@unibo.it}
\emailAdd{lauro.moscardini@unibo.it.}

\abstract{In the very last years, cosmological models where the properties of the dark components of the Universe --- dark matter and dark energy --- are accounted for by a single ``dark fluid'' have drawn increasing attention and interest. Amongst many proposals, Unified Dark Matter (UDM) cosmologies are promising candidates as effective theories. In these models, a scalar field with a non-canonical kinetic term in its Lagrangian mimics both the accelerated expansion of the Universe at late times and the clustering properties of the large-scale structure of the cosmos. However, UDM models also present peculiar behaviours, the most interesting one being the fact that the perturbations in the dark-matter component of the scalar field do have a non-negligible speed of sound. This gives rise to an effective Jeans scale for the Newtonian potential, below which the dark fluid does not cluster any more. This implies a growth of structures fairly different from that of the concordance \lcdm\ model. In this paper, we demonstrate that forthcoming large-scale surveys will be able to discriminate between viable UDM models and \lcdm\ to a good degree of accuracy. To this purpose, the planned Euclid satellite will be a powerful tool, since it will provide very accurate data on galaxy clustering and the weak lensing effect of cosmic shear. Finally, we also exploit the constraining power of the ongoing CMB Planck experiment. Although our approach is the most conservative, with the inclusion of only well-understood, linear dynamics, in the end we also show what could be done if some amount of non-linear information were included.}


\arxivnumber{}

\begin{document}

\maketitle

\section{Introduction}
The issue of finding a theoretically satisfying explanation for the so-called ``dark sector'' of the Universe has become one of the main challenges of contemporary cosmology. The current  concordance $\Lambda$ cold dark matter (\lcdm) paradigm seems to be a successful model of our cosmos, for its outstanding agreement with many different datasets. Nonetheless, there are still several open problems, one of the most compelling one being the presence of the two unknown dark-energy and dark-matter components in the Universe's total energy budget.

As a matter of fact, data coming from several cosmological r\'egimes and unrelated observables point towards the description of a Universe governed for $\sim75\%$ by a form of gravitationally-repulsive dark energy --- well described by a cosmological constant $\Lambda$ --- and for $\sim20\%$ by dark matter, i.e. a kind of matter which interacts (almost) only gravitationally and does not emit light. The main observational pieces of evidence favouring this scenario are the temperature anisotropy pattern of the Cosmic Microwave Background (CMB) radiation \citep{deBernardis:2000gy,Stompor:2001xf,Netterfield:2001yq,Rebolo:2004vp,Bennett:2003bz,Spergel:2003cb,Komatsu:2008hk}; the Hubble diagram of type Ia supernov\ae\ \citep{Perlmutter:1996ds,Riess:1998cb,Schmidt:1998ys,Perlmutter:1998np,Knop:2003iy,Tonry:2003zg,Riess:2004n,Astier:2005qq,WoodVasey:2007jb}; the clustering properties probing the cosmic Large-Scale Structure (LSS) \citep{Percival:2002gq,Pope:2004cc}; and the problem of the missing mass in the dynamics of galaxies and galaxy clusters \citep{Zwicky:1933gu,Zwicky:1937zza,Dodelson:2001ux,Hawkins:2002sg,Spergel:2006hy,Riess:2006fw}.

However, a comprehensive interpretation of the nature of the dark components is still far from being achieved. Moreover, there are several open issues. For instance, it has been shown that dark matter must be constituted by weakly-interacting massive particles rather than massive astrophysical compact-halo objects \citep{Tisserand:2006zx,Wyrzykowski:2009ep}. Nevertheless, no hint of such particles or their signatures have been detected so far, although the strong effort made by astroparticle and experimental physicists \citep{Bertone:2004pz,Fornengo:2006yy,Feng:2010gw}. For what concerns dark energy, the most puzzling problem is probably the $\sim120$ order-of-magnitude discrepancy between the cosmological-constant observed value and what estimated by quantum field theory. This is usually referred to as the ``cosmological-constant problem'' \citep{Weinberg:1988cp,Carroll:2000fy}. Eventually, we also lack the understanding of why these dark components do have comparable magnitudes today. This is the so-called ``coincidence problem.''

For all these reasons, in the last decades a large number of alternative cosmologies have been proposed. Amongst them, in this paper we focus on a specific class \citep{Bertacca:2010ct} of so-called quartessence or Unified Dark Matter (UDM) models \citep[e.g.][]{Kamenshchik:2001cp,Bilic:2001cg,Bento:2002ps,Balbi:2007mz,Quercellini:2007ht,Scherrer:2004au,Bertacca:2007ux,Chimento:2009nj,Piattella:2009kt,Bertacca:2010mt}. Here, the properties of a (classical) scalar field with a non-canonical kinetic term in its Lagrangian take the dynamics of both dark matter and dark energy into account. Albeit the background evolution of the Universe can be tailored in order to exactly reproduce \lcdm\ --- and therefore its agreement with ``geometrical'' datasets  ---, one of the principal peculiarities of these models is the arising of a sound speed for the perturbations of the dark-matter-like component of the scalar-field energy density. Such a sound speed induces an effective Jeans length and, as a result, matter perturbations cannot cluster on scales smaller than it.

Recently, these models have drawn an increasing interest. Bertacca et al. \citep{Bertacca:2008uf} have demonstrated how to construct a viable class of UDM model with only one extra parameter compared to \lcdm. These models could exhibit a sound speed small enough to allow the clustering of the cosmic structures we see today without being plagued by an integrated Sachs-Wolfe effect too large to be compatible with CMB data \citep{Bertacca:2007cv}. Some tests and constraints have already been posed to this class of UDM models \citep{Camera:2009uz,Camera:2010wm,Bertacca:2011in}. In this paper, we aim to explore the potential of near-future wide-field surveys in constraining these UDM models. Indeed, their interesting behaviour in the growth of matter over-densities will be clearly detectable by a large-scale survey such as Euclid\footnote{http://www.euclid-ec.org} \citep{EditorialTeam:2011mu} and a CMB probe like Planck\footnote{http://www.esa.int/planck} \citep{:2006uk}.

The paper is structured as follows. In Section~\ref{sec:udm}, we briefly outline the most important theoretical aspects of the class of UDM models we analyse here. Section~\ref{sec:observables} is devoted to describe the cosmological observables used. Specifically, we exploit the power of the Baryon Acoustic Oscillation (BAO) signature in the galaxy-galaxy power spectrum (Section~\ref{ssec:bao}); information about the large-scale clustering and its time-dependent evolution encoded in the weak lensing effect of cosmic shear (Section~\ref{ssec:shear}); and we also make use of the capability of CMB experiments in constraining cosmological parameters and lifting many degeneracies amongst them (Section~\ref{ssec:cmb}). The characteristics of the large-scale surveys we adopt are presented in Section~\ref{sec:surveys}. Our results are discussed in Section~\ref{sec:results}, whilst in Section~\ref{sec:conclusions} we draw the main conclusions.

\section{Unified Dark Matter Models}\label{sec:udm}
UDM models are a class of scalar-field theories. Here, a scalar field $\varphi(t)$ accounts for both dark matter and $\Lambda$ thanks to a non-canonical kinetic term in its Lagrangian. Hereafter, we choose a Lagrangian of the form \citep{Bertacca:2008uf,Bertacca:2010ct}
\begin{equation}
\mathscr L_\mathrm{UDM}=f(\varphi)g(X)-V(\varphi)\label{L_phi},
\end{equation}
with a Born-Infeld type kinetic term $g(X)=-\sqrt{1-2X/\Lambda}$ \citep{Born:1934gh}. Such a kinetic term can be thought as a field theory generalisation of the Lagrangian of a relativistic particle \citep{Padmanabhan:2002sh,Abramo:2003cp,Abramo:2004ji}. It was also proposed in connection with string theory, since it seems to represent a low-energy effective theory of D-branes and open strings, and has been conjectured to play a r\^ole in cosmology \citep{Sen:2002nu,Sen:2002in,Sen:2002vv,Padmanabhan:2002sh}. By using the equation of motion of the homogeneous scalar field $\varphi(t)$ and by imposing that the scalar field Lagrangian is constant along the classical trajectories, i.e. $p_\mathrm{UDM}=-\rho_\Lambda$, we can obtain the following expressions for the potentials
\begin{align}
f(\varphi)&=\frac{\Lambda \cinf}{1-{\cinf}^2}\frac{\cosh(\xi\varphi)}{\sinh(\xi\varphi)\left[1+\left(1-{\cinf}^2\right)\sinh^2(\xi\varphi)\right]},\\V(\varphi)&=\frac{\Lambda}{1- {\cinf}^2}\frac{\left(1-{\cinf}^2\right)^2\sinh^2\left(\xi\varphi\right)+2(1-{\cinf}^2)-1}{1+\left(1-{\cinf}^2\right)\sinh^2\left(\xi\varphi\right)} \;,
\end{align}
with $\xi=\sqrt{3/[4(1-{\cinf}^2)]}$ and $\cinf$ a free parameter.

\subsection{Background Evolution}
We choose the Universe's metric according to a Friedmann-Lema\^itre-Robertson-Walker line element with $a(t)$ the scale factor as a function of the cosmic time $t$. We use units such that $c=1$ and signature $\{-,+,+,+\}$. The energy density of the UDM scalar field considered here presents two terms,
\begin{equation}
\rho_\mathrm{UDM}(t)=\rho_\mathrm{DM}(t)+\rho_\Lambda,
\end{equation}
where $\rho_\mathrm{DM}$ behaves like a dark matter component ($\rho_\mathrm{DM}\propto a^{-3}$) and $\rho_\Lambda$ like a cosmological constant component ($\rho_\Lambda=\mathrm{const.}$) Consequently, in these UDM models the Hubble parameter $H\equiv\de\ln a/\de t$ is the same as in \lcdm, namely
\begin{equation}
H(z)=\ho\sqrt{\om\left(1+z\right)^3+\ol},\label{eq:hubble}
\end{equation}
with $\ho=100\,h\,\mathrm{km\,s^{-1}\,Mpc^{-1}}$ the Hubble constant, $\om=\odm+\ob$ the total matter fraction, given $\odm$ and $\ob$ the dark-matter and baryon densities in units of the critical density, and $z=1/a-1$ the cosmological redshift.

\subsection{Growth of Cosmic Structures}
Differences compared to \lcdm\ arise in the growth of cosmic structures. The linearised perturbed metric in the Newtonian gauge reads
\begin{equation}
\de s^2=-(1+2\Phi)\de t^2+a^2(t)(1+2\Psi)\de\mathbf x^2,\label{flrw}
\end{equation}
with $\Phi$ and $\Psi$ the two Bardeen's potentials \citep{Bardeen:1980kt}. For the two metric perturbations $\Phi=-\Psi$ still holds, because the scalar field does not present any anisotropic stress. From Einstein's perturbed field equations one obtains the evolution equation for the Newtonian potential $\Phi$ \citep{Mukhanov:2005sc},
\begin{equation}
\Phi''+3\left(1+3{c_s}^2\right)\mathcal H\Phi'-{c_s}^2\nabla^2\Phi+\left[2\mathcal H'+\left(1+3{c_s}^2\right)\mathcal H^2\right]\Phi=0,\label{eq:phi}
\end{equation}
where a prime denotes a derivative with respect to the conformal time $\de\tau=\de t/a$, and $\mathcal H\equiv\de\ln a/\de\tau$ is the conformal Hubble parameter. Here, $c_s(a)$ is the effective speed of sound. For the present Lagrangian we have \citep{Bertacca:2008uf}
\begin{equation}
{c_s}^2(a)=\frac{{\ol \cinf}^2}{\ol+(1-{\cinf}^2)\odm a^{-3}}\label{c_s-udm},
\end{equation}
and it is easy to see that the extra parameter $\cinf$ represents the value of the speed of sound when $a\rightarrow\infty$. Moreover, when $a\to0$, $c_s\to0$.

To summarise, in UDM models the fluid which leads to the observed accelerated expansion is also the one which clusters and produces the cosmic structures we see today. Thus, from equality to the present epoch, the energy density of the Universe is dominated by a single ``dark fluid,'' and the gravitational potential evolution is therefore determined by the background and perturbation evolution of this fluid alone. As a result, the general trend is that the possible appearance of a sound speed significantly different from zero at late times corresponds to the appearance of a Jeans length $\lambda_J(a)\propto c_s(a)$ \citep{Bertacca:2007cv} below which the dark fluid does not cluster any more, causing a strong evolution in time of the gravitational potential \citep{Camera:2009uz,Camera:2010wm}. Since in UDM models gravity is still described by General Relativity, the growth of cosmic structures and matter over-densities can be studied by using the canonical Poisson equation
\begin{equation}
\nabla^2\Phi(t,\mathbf x)=\frac{3}{2}\om{\ho}^2\frac{\delta(t,\mathbf x)}{a(t)}\label{eq:poisson-udm},
\end{equation}
which relates the matter density contrast $\delta(t,\mathbf x)$ to the Newtonian potential $\Phi(t,\mathbf x)$. The latter is obtained from eq.~\eqref{eq:phi}.

\section{Cosmological Observables}\label{sec:observables}
We adopt the Fisher matrix formalism \citep{Fisher:1935,Jungman:1995bz,Tegmark:1996bz} to make predictions on UDM models from future large-scale surveys. Specifically, we describe our approach based on the Fisher matrices for a LSS survey such as Euclid and a CMB experiment like Planck. For the former, we exploit both its spectroscopic measurements --- to study galaxy clustering --- and its photometric data --- to analyse the cosmic shear effect.

The Fisher matrix is the expectation value of the second derivative of the natural logarithm of the likelihood function, $L$, with respect to the cosmological parameters $\{\vartheta_\alpha\}$, i.e.
\begin{equation}
\mathbf F_{\alpha\beta}=-\left\langle\frac{\partial^2\ln L}{\partial\vartheta_\alpha\partial\vartheta_\beta}\right\rangle\label{fisherm}
\end{equation}
and the marginal error on parameter $\vartheta_\alpha$ is $\sqrt{\left(\mathbf F^{-1}\right)_{\alpha\alpha}}$. The total Fisher matrix that we scrutinise here is
\begin{equation}
\mathbf F=\mathbf F^\mathrm{LSS}+\mathbf F^\mathrm{CMB},
\end{equation}
where $\mathbf F^\mathrm{LSS}=\mathbf F^\mathrm{BAO}+\mathbf F^\gamma$ and $\mathbf F^\mathrm{CMB}=\mathbf F^T+\mathbf F^E+\mathbf F^{TE}$, being $\mathbf F^\mathrm{BAO}$, $\mathbf F^\gamma$, $\mathbf F^T$, $\mathbf F^E$ and $\mathbf F^{TE}$ the contributions from BAO, cosmic shear tomography and CMB temperature, $E$-polarisation and their cross-correlation, respectively.

Finally, another important quantity that we calculate is the correlation between the parameter pair $\vartheta_\alpha$-$\vartheta_\beta$, which is defined as
\begin{equation}
r_{\vartheta_\alpha-\vartheta_\beta}=\frac{\left(\mathbf F^{-1}\right)_{\alpha\beta}}{\sqrt{\left(\mathbf F^{-1}\right)_{\alpha\alpha}\left(\mathbf F^{-1}\right)_{\beta\beta}}}.\label{eq:correlation}
\end{equation}
This quantity tells us whether the two parameters are completely uncorrelated ($\left|r_{\vartheta_\alpha-\vartheta_\beta}\right|=0$) or completely degenerate ($\left|r_{\vartheta_\alpha-\vartheta_\beta}\right|=1$), with all the hues in between.

\subsection{Large-Scale Structure}
\begin{figure}
\centering
\includegraphics[width=0.9\textwidth]{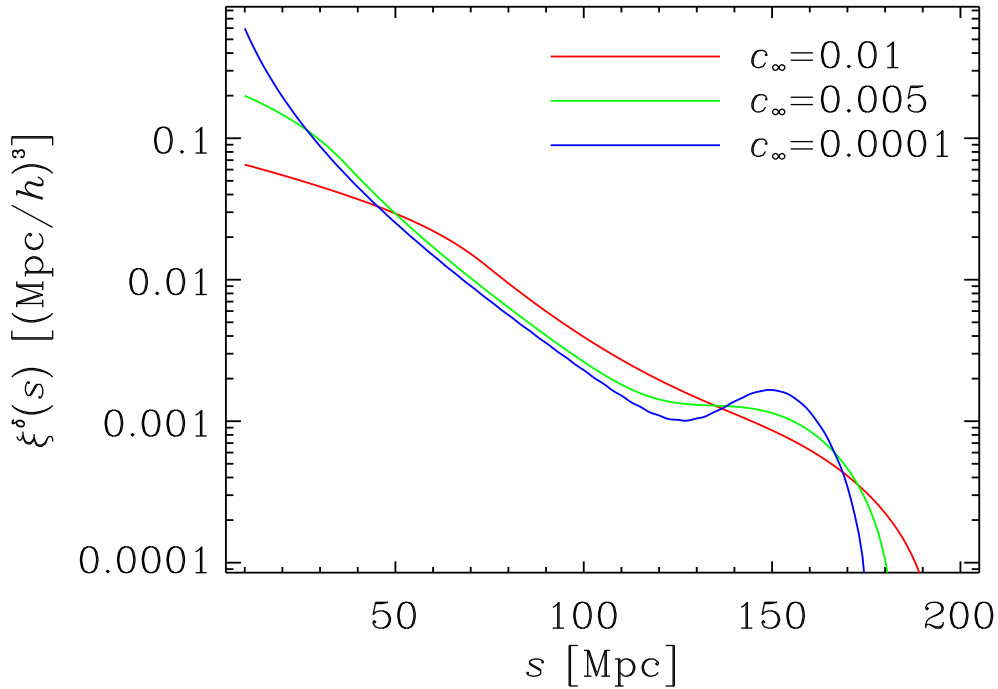}
\caption{Present-day two-point matter correlation function $\xi^\delta(s)$ as a function of comoving separation $s$ for UDM models with $\cinf=0.01$ (red), $0.005$ (green) and $0.0001$ (blue).}\label{fig:correlation-function-UDM}
\end{figure}

The main observable we use concerning the Universe's LSS is the matter power spectrum $P^\delta(k,z)$. It is defined as the Fourier transform of the two-point correlation function, which at a given time $t$ reads
\begin{equation}
\xi^\delta(s)=\langle\delta(\mathbf x)\delta^\star(\mathbf y)\rangle,\label{eq:x^d}
\end{equation}
with $s=|\mathbf x-\mathbf y|$ the separation between the two positions. Figure~\ref{fig:correlation-function-UDM} shows $\xi^\delta(s)$ at $z=0$, as expected in UDM models with different $\cinf$'s. Specifically, the fiducial value $0.0001$ (blue curve) as well as $0.005$ (green curve) and $0.01$ (red curve). Our fiducial UDM model clearly shows the BAO peak at $\sim150\,h^{-1}\,\mathrm{Mpc}$. Contrarily, the larger is the speed of sound, the more suppressed is the perturbation clustering on small scales --- as expected (see Section~\ref{sec:udm}). Moreover, we can also notice that a large value of $\cinf$ not only inhibits the growth of structures below the Jeans length $\lambda_J$, but also acts by sweeping the BAO signatures out. This is easy to understand: we shall be able to observe the BAO peak of matter perturbations only if the BAO characteristic scale is larger than the Jeans length. Otherwise, the suppression of the Newtonian potential due to the hydrodynamics of the scalar field drags any over-density away.

The UDM peculiar growth of density perturbations, which comes from eqs.\eqref{eq:phi}-\eqref{eq:poisson-udm}, can be accounted for by a time- and scale-dependent term in the power spectrum \citep{Bertacca:2011in}. Here, we use the analytical approximation of Piattella and Bertacca \citep{Piattella:2011fv}, which has been shown to depart from the actual solution by less than $0.5\%$. Specifically, it reads
\begin{equation}
T_\mathrm{UDM}(k,z)=2^{5/8}\Gamma\left[k\mathcal A(z)\right]^{-5/8}J_{5/8}\left[k\mathcal A(z)\right],
\end{equation}
where $\Gamma$ is Euler's gamma function, $J_\nu$ is Bessel's function of order $\nu$ and
\begin{equation}
\mathcal A\left[z(a)\right]=\int\!\!\de a\,\frac{c_s(a)}{a^2H(a)}.
\end{equation}
Hence, in the linear r\'egime of perturbations, the power spectrum of the clustering component of the UDM scalar field is
\begin{equation}
P^\delta(k,z)=\frac{8\pi^2k_0A_s}{25\ho^4\om^2}\left(\frac{k}{k_0}\right)^{n_s}\left[T_\mathrm{UDM}(k,z)T(k)\frac{D_+(z)}{D_+(z=0)}\right]^2,\label{eq:P_k}
\end{equation}
with $n_s$ the scalar spectral index of the primordial power spectrum, $A_s$ the dimensionless amplitude of the primordial curvature perturbations at the pivot scale $k_0=0.002\,\mathrm{Mpc}^{-1}$ and $T(k)$ the matter transfer function, which describes the evolution of perturbations through the epochs of horizon crossing and radiation-matter transition. Here, $D_+(z)$ is the growth factor, and it is defined as time-dependent part of the density contrast. The present-day matter power spectrum is usually referred to as $P(k)\equiv P^\delta(k,z=0)$.

\subsubsection{Baryon Acoustic Oscillations}\label{ssec:bao}
In this section we investigate the galaxy power spectrum $P^g(k,z)$. We utilise the so-called $P(k)$-method marginalised over growth information, which exploits only the shape and the BAO positions of the matter power spectrum. This means that we marginalise over the amplitude and redshift-space distortions. This method allows us to estimate the cosmological parameters which characterise the underlying fiducial cosmology from the galaxy catalogue. It has already been applied to the galaxy power spectrum in a number of works \citep[see e.g.][]{Seo:2003pu,Wang:2006qt,Wang:2007ht,Wang:2008zh,Wang:2010gq,Carbone:2010ik,Carbone:2011bx,Carbone:2011by,DiPorto:2012ey,Amendola:2011ie,DiPorto:2011jr}. Here, we present the main formul\ae\ describing this approach.

When we include redshift-space distortions, the galaxy power spectrum reads
\begin{equation}
P^g(k_\perp,k_\parallel;z)={b_g}^2(z)\left[1+\beta(k,z)\frac{{k_\parallel}^2}{{k_\perp}^2+{k_\parallel}^2}\right]^2P^\delta(k,z)e^{-k^2\mu^2{\sigma_\chi}^2},\label{eq:Pg}
\end{equation}
where $k_\perp$ and $k_\parallel$ are the wave-numbers across and along
the line of sight, $b_g(z)$ is the redshift-dependent galaxy bias and $\beta(k,z)$ is the linear redshift-space distortion parameter \citep{Kaiser:1987qv}. We also add the damping factor $e^{-k^2\mu^2{\sigma_\chi}^2}$ to take redshift uncertainties into account, with $\mu=\widehat{\mathbf k}\cdot\widehat{\mathbf n}$ the cosine between $\mathbf k$ and the line-of-sight direction $\widehat{\mathbf n}$, $\sigma_\chi=(\partial\chi/\partial
z)\sigma_z$, $\de\chi=\de H(z)/\de z$ being the differential radial comoving distance and $\sigma_z$ the error on redshift measurements \citep{Wang:2009gt,Seo:2003pu}.

However, the observed galaxy power spectrum actually is
\begin{equation}
P^g_\mathrm{obs}(k_{\mathrm{ref}\perp},k_{\mathrm{ref}\parallel};z)=\frac {\left[{d_A}^2(z)\right]_\mathrm{ref}H(z)}{{d_A}^2(z)\left[H(z)\right]_\mathrm{ref}}P^g(k_{\mathrm{ref}\perp},k_{\mathrm{ref}\parallel};z)+P_\mathrm{shot},\label{eq:Pobs_here}
\end{equation}
where the $P^g(k_{\mathrm{ref}\perp},k_{\mathrm{ref}\parallel};z)$ pre-factor encapsulates the geometrical distortions due to the  Alcock-Paczynski
effect \citep{Seo:2003pu,Ballinger:1996cd,Marulli:2012na}, with $d_A(z)=\chi(z)/(1+z)$ the angular diameter distance. Values in the reference cosmology are distinguished by the subscript `ref,' whilst those in the true cosmology have no subscript. Regarding the wave-numbers, we have $k_{\mathrm{ref}\perp}=k_\perp\chi/\chi_\mathrm{ref}$ and $k_{\mathrm{ref}\parallel}=k_\parallel H_\mathrm{ref}/H$. Finally, $P_\mathrm{shot}$ is the unknown white shot noise that remains even after the conventional shot noise of inverse number density has been subtracted \citep{Seo:2003pu}. It could arise from clustering bias even on large scales due to local bias \citep{Seljak:2000gq}.

We divide the survey volume $V_\mathrm{survey}$ in redshift shells with size $\Delta z=0.1$, centred at redshift $z_i$, and choose the following set of
parameters 
\begin{equation}
\left\{H(z_i),\,d_A(z_i),\,\overline{D_+}(z_i),\,\beta(k,z_{i}),\,P_\mathrm{shot}^i,\,\omega_m,\,\omega_b,\,\cinf,\,n_s,h\right\}
\end{equation}
to describe $P^{g}_\mathrm{obs}(k_{\mathrm{ref}\perp},k_{\mathrm{ref}\parallel};z)$. Here, $\omega_m=\Omega_m h^2$ and $\omega_b=\Omega_b h^2$ are the so-called physical matter and baryon fractions. Finally, since the growth factor $D_+(z)$, the bias $b_g(z)$ and the power spectrum normalisation $A_s$ are completely degenerate, we introduce the quantity $\overline{D_+}(z_i)=\sqrt{\mathcal P_0}b_g(z_i)D_+(z_i)/D_+(z=0)$ \citep{Wang:2008zh,Marulli:2011jk,Marulli:2011he}, with $\mathcal P_0$ a handle for the power spectrum normalisation.

It has been shown that the Fisher matrix associated to the observed galaxy power spectrum can be well approximated by \citep{Tegmark:1996bz,Tegmark:1997rp} 
\begin{equation}
\mathbf F_{ij}^\mathrm{BAO}=\int_{-1}^1\!\!\de\mu\,\int_{k_\mathrm{min}}^{k_\mathrm{max}}\frac{\de k}{4\pi^2}\,k^{2}\frac{\partial\ln P^{g}_\mathrm{obs}(k,\mu)}{\partial\vartheta_i}\frac{\partial\ln P^{g}_\mathrm{obs}(k,\mu)}{\partial\vartheta_j}V_\mathrm{eff}(k,\mu),\label{Fisher}                 
\end{equation}
where the derivatives are evaluated at the parameter values $\vartheta_i$ of the fiducial model and $V_\mathrm{eff}$ is the effective volume of the survey, namely
\begin{equation}
V_\mathrm{eff}(k,\mu)=\left[\frac{N_gP^{g}(k,\mu)}{N_gP^{g}(k,\mu)+1}\right]^2V_\mathrm{survey},\label{V_eff} 
\end{equation}
$N_g$ being the mean galaxy number density. We do not include information from the amplitude $\overline{D_+}(z_i)$ and the redshift space distortions $\beta(k,z_i)$, we thus marginalise over these parameters and also over $P_\mathrm{shot}^i$. Then, we project $\{H(z_i)$, $d_A(z_i)$, $\omega_m$, $\omega_b$, $\cinf$, $n_s$, $h\}$ into the final set
\begin{equation}
\left\{\om,\,\ob,\,h,\,\cinf,\,n_s\right\}.\label{q_set}
\end{equation}
The transformation from one set of parameters to another is given by
\begin{equation}
\mathbf F^\mathrm{BAO}_{\alpha \beta}=\mathbf J^{i}_{\alpha}\mathbf F^\mathrm{BAO}_{ij}\mathbf J^{j}_{\beta},\label{eq:Fisherconv}
\end{equation}
where $\mathbf J^{i}_{\alpha}$ is the Jacobian matrix of the coordinate transformation in the parameter space.

\subsubsection{Cosmic Shear}\label{ssec:shear}
Weak gravitational lensing is responsible for the shearing and magnification of the images of high-redshift sources due to the presence of intervening matter. The distortions are due to fluctuations in the gravitational potential, and are directly related to the distribution of matter and to the geometry and dynamics of the Universe. In particular, the distortions occurring to a background image can be decomposed into a convergence $\kappa$ and a (complex) shear $\gamma=\gamma_1+i\gamma_2$, which are the entries of the distortion matrix
\begin{equation}
\mathbf D=\left(\begin{array}{cc}
\kappa+\gamma_1&\gamma_2\\
\gamma_2&\kappa -\gamma_1
\end{array}\right).
\end{equation}
The distortion matrix is directly related to background and perturbed cosmological quantities via \citep{Kaiser:1996tp,Bartelmann:1999yn}
\begin{equation}
\mathbf D_{ij}=\int\!\!\de\chi\,\chi W(\chi)\delta_{,ij}(\widehat{\mathbf n},\chi)\label{phi,ij},
\end{equation}
with commas denoting derivatives with respect to directions perpendicular to the line of sight $\widehat{\mathbf n}$. Here,\begin{equation}
W(\chi)=\frac{3}{2}\ho^2\om\left[1+z(\chi)\right]\chi\int_\chi^\infty\de\chi'\,\frac{\chi'-\chi}{\chi'}\frac{\de N_g}{\de\chi'}(\chi')\label{W(z)}
\end{equation}
is the weight function of weak lensing, and $\de N_g/\de z\left[\chi(z)\right]$ represents the redshift distribution of the sources, normalised such that $\int\!\!\de N_g=1$.

In the flat-sky approximation, we expand the shear $\gamma(\widehat{\mathbf n})$ in its Fourier modes and the two-dimensional angular power spectrum $C^\gamma(\ell)$ is given by
\begin{equation}
\langle\gamma(\vec{\ell})\gamma^\ast(\vec{\ell'})\rangle={(2\pi)}^2\delta_D(\vec{\ell}-\vec{\ell'})C^\gamma(\ell).
\end{equation}
In the case where one has distance information for individual sources, we can use this information for statistical studies. A natural course of action is to divide the survey into slices at different distances, and perform a study of the shear pattern on each slice \citep{Hu:1999ek}. In order to use the information effectively, it is necessary to look at cross-correlations of the shear fields in the slices, as well as correlations within each slice. This procedure is usually referred to as tomography. By doing so, the two-dimensional tomographic shear matrix becomes
\begin{equation}
\mathbf C^\gamma_{ij}(\ell)=\int\!\!\de\chi\,\frac{W_i(\chi)W_j(\chi)}{\chi^2}P^\delta\!\left(\frac{\ell}{\chi},\chi\right),\label{eq:tomography}
\end{equation}
where $W_i(\chi)$ is the weight function \eqref{W(z)} related to the $i$-th bin and we have introduced Limber's approximation, in which the only Fourier modes that contribute to the integral are those with $\ell=k\chi$.

For cosmic shear tomography, for a square patch of the sky, the Fourier transform leads to uncorrelated modes, provided the modes are separated by $2\pi/\Theta_\mathrm{rad}$ where $\Theta_\mathrm{rad}$ is the side of the square in radians. Then, the Fisher matrix is simply the sum of the Fisher matrices of each $\ell$ mode,
\begin{equation}
\mathbf F_{\alpha\beta}^\gamma=\sum_\ell\frac{2\ell+1}{2}f_\mathrm{sky}^\mathrm{Euclid}\frac{\partial\mathbf C_{ij}^\gamma(\ell)}{\partial\vartheta_\alpha}\left[\mathbf C^{\gamma,\gamma}_\ell\right]_{jk}^{-1}\frac{\partial\mathbf C_{kl}^\gamma(\ell)}{\partial\vartheta_\beta},\label{eq:Fisher}
\end{equation}
where $f_\mathrm{sky}^\mathrm{Euclid}$ is the fraction of the sky covered by the survey under analysis and $\left[\mathbf C^{\mu,\mu}_\ell\right]_{ij}$ is the covariance matrix for a given $\ell$ mode and the $i$-$j$ bin pair, i.e.
\begin{equation}
\left[\mathbf C^{\gamma,\gamma}_\ell\right]_{ij}=\left[\mathbf C^{\gamma}_{ij}(\ell)+\frac{{\langle\gamma_\mathrm{int}}^{2}\rangle}{N_g}\delta_{ij}\right]^2,
\end{equation}
with $\langle{\gamma_\mathrm{int}}^2\rangle^{0.5}\simeq0.4$ the galaxy-intrinsic shear rms in one component and $\delta_{ij}$ the Kronecker symbol.

\begin{figure}
\centering
\includegraphics[width=0.9\textwidth]{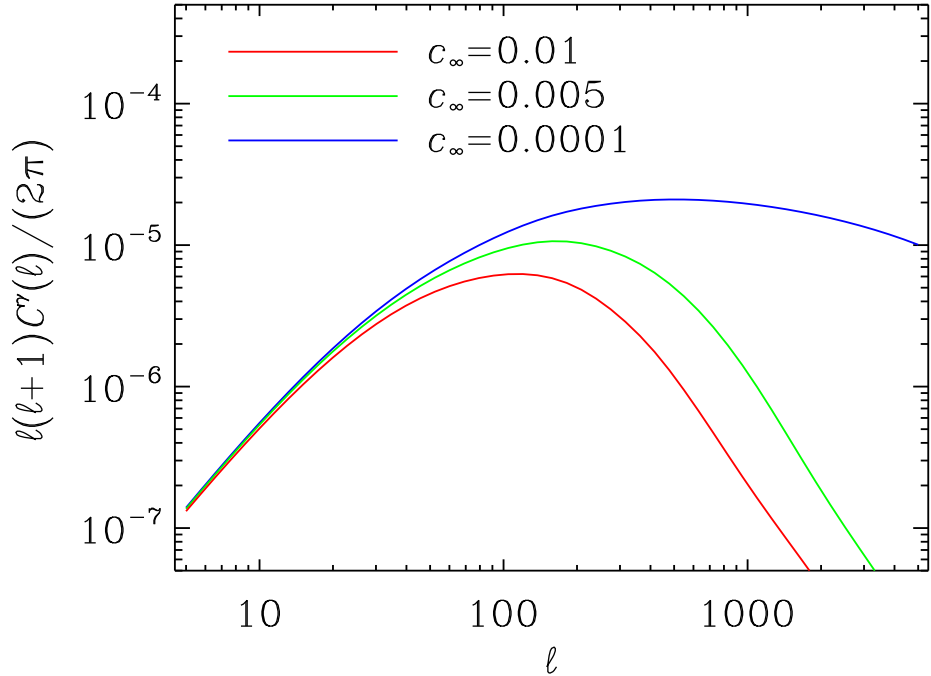}
\caption{Cosmic-shear power spectrum as a function of the angular scale $\ell$ for UDM models with $\cinf=0.01$ (red), $0.005$ (green) and $0.0001$ (blue).}\label{fig:shear-UDM}
\end{figure}

Figure~\ref{fig:shear-UDM} shows the linear UDM shear power spectrum $\ell(\ell+1)C^\gamma(\ell)/(2\pi)$ for sound speeds with $\cinf=0.0001$ (blue curve), $0.005$ (green curve) and $0.01$ (red curve). As in Figure~\ref{fig:correlation-function-UDM}, it is easy to see the effect due to an increasing $\cinf$, which suppresses the Newtonian potential on scales smaller than the inverse of the Jeans length. This reflects on the shear power spectrum through Limber's equality $k=\ell/\chi$.

\subsection{Cosmic Microwave Background}\label{ssec:cmb}
Here, we make use of the Planck-mission parameter constraints as CMB priors by estimating the cosmological parameter errors via measurements of the temperature and polarisation power spectra. The main observables are the angular power spectrum of the CMB temperature anisotropies $C^T_\ell$ and its $E$-polarisation $C^E_\ell$. We do not include any $B$-mode in our forecasts and assume no tensor mode contribution to the power spectra.

The Fisher matrix for CMB power spectrum is given by \citep{Zaldarriaga:1996xe,Zaldarriaga:1997ch}
\begin{equation}
\mathbf F_{ij}^\mathrm{CMB}=\sum_{\ell}\frac{2\ell+1}{2}f_\mathrm{sky}^\mathrm{Planck}\sum_{X,Y}\frac{\partial C^X_\ell}{\partial\vartheta_i}\left[\mathbf C_\ell^{X,Y}\right]^{-1}\frac{\partial C^Y_\ell}{\partial\vartheta_{j}},\label{eqn:cmbfisher}
\end{equation}
where $f^\mathrm{Planck}_\mathrm{sky}$ is the sky fraction covered by Planck and $C^X_\ell$ is the harmonic power spectrum for the temperature-temperature ($X\equiv T$), temperature-$E$ polarisation ($X\equiv TE$) and the $E$ polarisation-$E$ polarisation ($X\equiv E$) signal. The covariance matrix $\mathbf C_\ell^{X,Y}$ of the errors for the various power spectra is given by the fourth moment of the distribution, which under Gaussian assumptions is entirely given in terms of the $C^X_\ell$ with 
\begin{align}
\mathbf C_\ell^{T,T}&=\left(C^T_\ell+W_T^{-1}B_\ell^{-2}\right)^2, \\
\mathbf C_\ell^{E,E}&=\left(C^E_\ell+W_P^{-1}B_\ell^{-2}\right)^2,  \\
\mathbf C_\ell^{TE,TE}&=\left[\left(C^{TE}_\ell\right)^2+\left(C^T_\ell+W_T^{-1}B_\ell^{-2}\right)\left(C^E_\ell+W_P^{-1}B_\ell^{-2}\right)\right],\\
\mathbf C_\ell^{T,E}&=\left(C^{TE}_\ell\right)^2,  \\
\mathbf C_\ell^{T,TE}&=C^{TE}_\ell\left(C^{T}_\ell+W_T^{-1}B_\ell^{-2}\right), \\
\mathbf C_\ell^{E,TE}&=C^{TE}_\ell\left(C^{E}_\ell+W_P^{-1}B_\ell^{-2}\right),
\end{align}
where $W_{T,P}=\sum_c W^c_{T,P}$, $W^c_{T,P}=(\sigma^c_{T,P}\theta^c_\mathrm{fwhm})^{-2}$ is the weight per solid angle for temperature and polarisation respectively, with a 1-$\sigma$ sensitivity per pixel of $\sigma^c_{T,P}$ and a beam of $\theta^c_\mathrm{fwhm}$ extent, for each frequency channel $c$. The beam window function is given in terms of the full width half maximum (fwhm) beam width by $B_{\ell}^2 =\sum_c (B^c_{\ell})^2 W^c_{T,P}/W_{T,P}$, where $(B^c_\ell)^2= \exp\left[-\ell(\ell+1)/(\ell^c_s)^2\right]$, $\ell^c_s=2\sqrt{2\ln2}/\theta^c_\mathrm{fwhm}$ \citep{Bond:1997wr}.

\section{Large-Scale Surveys}\label{sec:surveys}
In this section, we briefly present the large-scale surveys we use in our analyses.

\subsection{Euclid Satellite}
As a LSS survey, we adopt a Euclid-like experiment for both spectroscopic galaxies measurements and photometric cosmic-shear data.

\subsubsection{Galaxy Survey}
\begin{figure}
\centering
\includegraphics[width=0.5\textwidth]{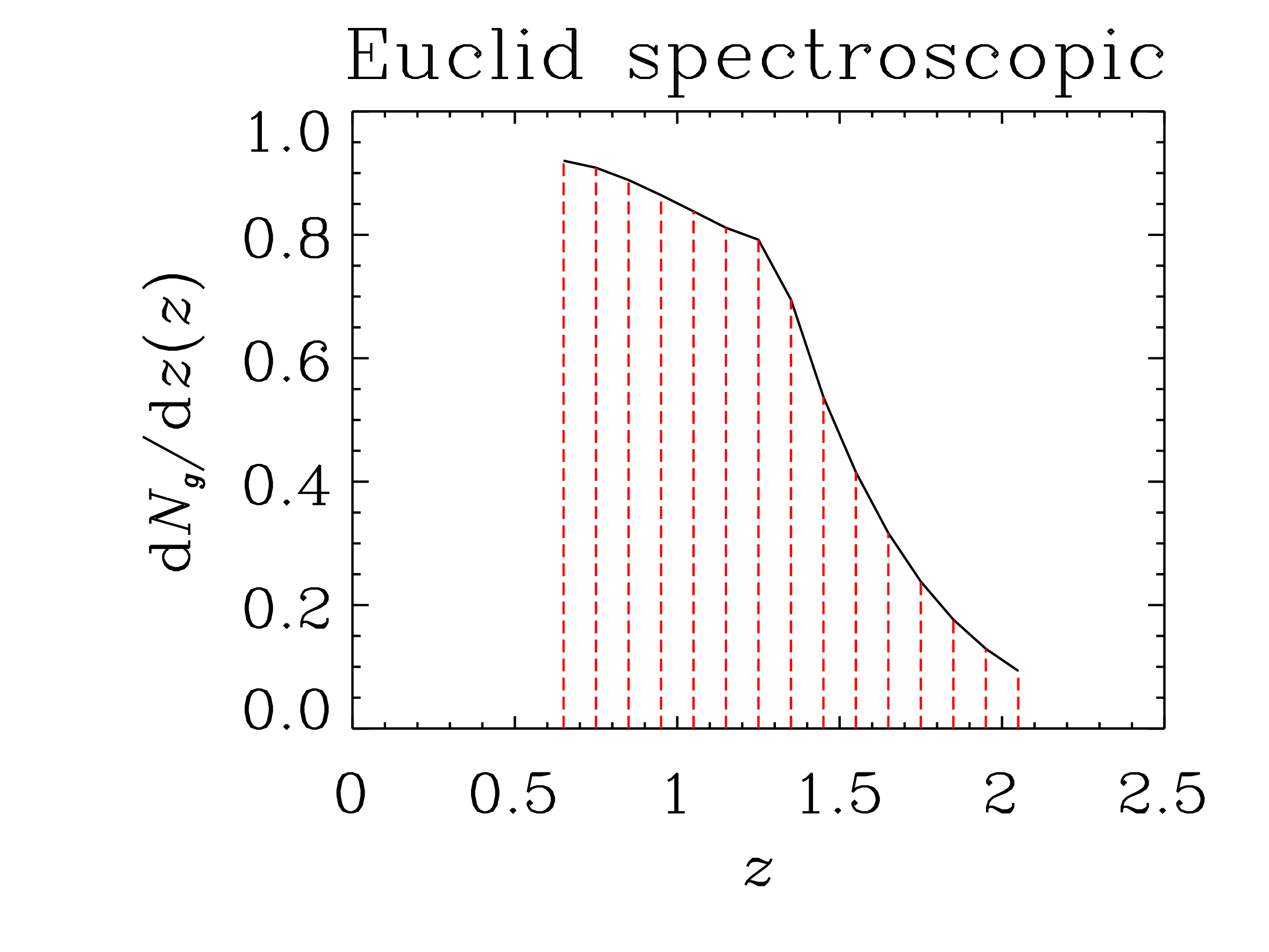}\includegraphics[width=0.5\textwidth]{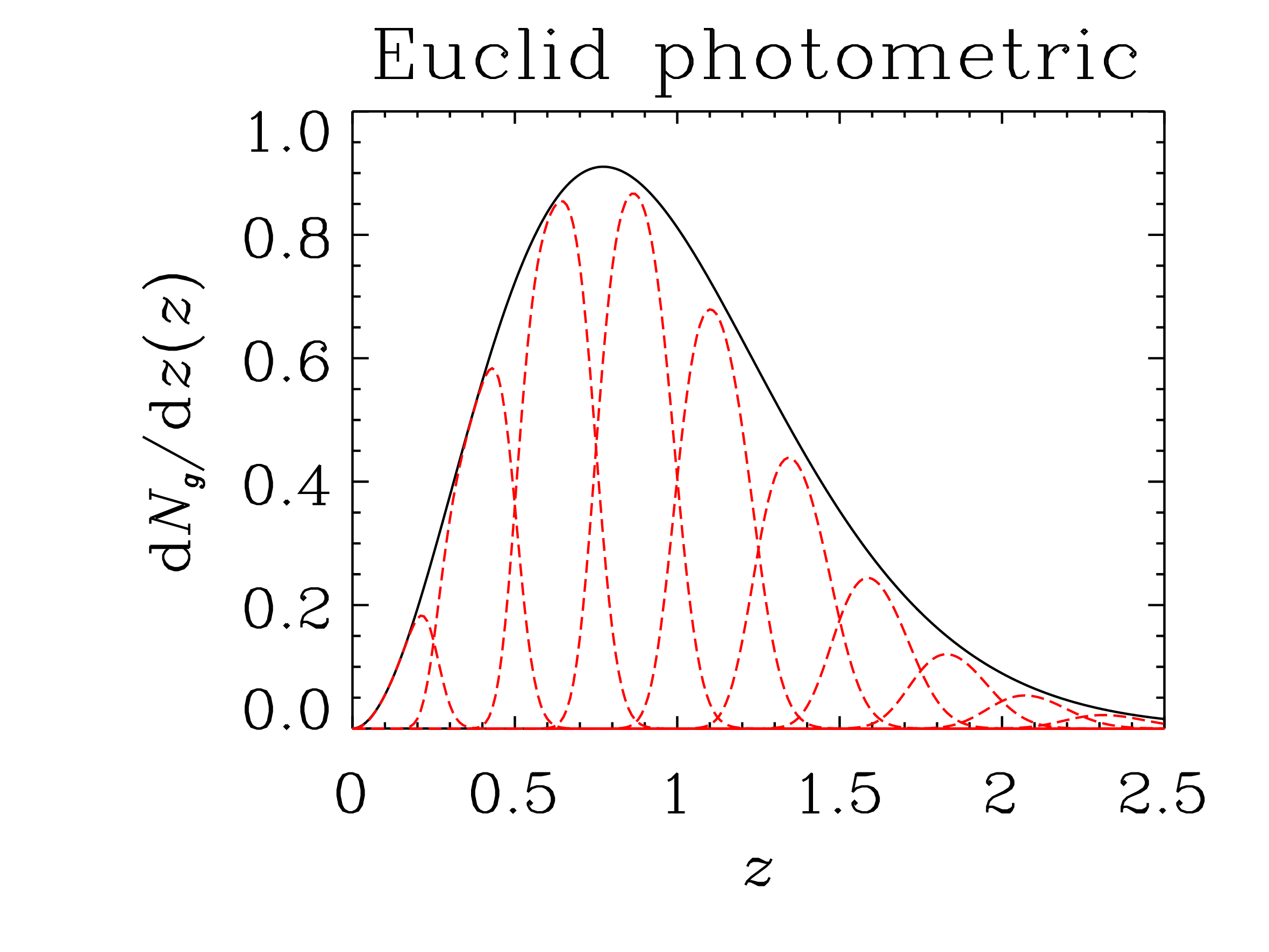}
\caption{\textit{Left panel:} The normalised Euclid-like redshift distributions of sources for the galaxy survey (solid, black) and the fourteen spectroscopic bins used (dashed, red). \textit{Right panel:} The normalised Euclid-like redshift distribution of sources for the cosmic-shear imaging survey (solid, black) and the ten photometric bins used (dashed, red).}\label{fig:dNdz-spectro_photo}
\end{figure}

For the spectroscopic galaxy survey, we adopt the empirical redshift distribution of H$\alpha$ emission line galaxies derived by \citep{2010MNRAS.402.1330G} from observed H$\alpha$ luminosity functions, and the bias function derived by \citep{2010MNRAS.405.1006O} using a galaxy formation simulation. In particular, we choose a flux limit of $4\cdot10^{-16}\,\mathrm{erg\,s^{-1}\,cm^{-1}}$, and a redshift success rate of $0.5$. We use the Euclid Red Book \citep{EditorialTeam:2011mu} specifications for the survey area 15,000 deg$^2$, the redshift accuracy $\sigma_z^\mathrm{sp}/(1+z)\le0.001$, and a redshift range $0.65\leq z\leq2.05$. With such an accuracy, the optimised subdivision in redshift bins has been found that shown in the left panel of Figure~\ref{fig:dNdz-spectro_photo}.

\subsubsection{Cosmic-Shear Survey}
We compute our results for a $15,000\,\mathrm{deg}^2$ cosmic-shear experiment. The source distribution over redshifts has the form \citep{1994MNRAS.270..245S}
\begin{equation}
\frac{\de N_g}{\de z}(z)\propto z^2e^{-\left(\frac{z}{z_0}\right)^{1.5}},\label{eq:n_z-Euclid}
\end{equation}
where $z_0=z_m/1.4$, and $z_m=0.9$ is the median redshift of the survey. The number density of the sources, with estimated photometric redshift and shape, is $30$ per square arcminute. To perform the cosmic-shear tomography explained in Section~\ref{ssec:shear}, we divide the redshift distribution of sources into ten redshift bins. However, the Euclid imaging survey will only provide photometric-redshift measurements, which are known to be less accurate than those obtained from spectroscopy. The scatter between the true redshift and the photometric estimate is assumed to be of order $3\%$ and scale linearly with $z$, that is to say $\sigma_z^\mathrm{ph}=0.03(1+z)$. The right panel of Figure~\ref{fig:dNdz-spectro_photo} illustrates the total $\de N_g/\de z$ (solid, black) and the ten photometric-redshift bins we use (dashed, red).

\subsection{Planck Satellite}
Our fiducial CMB experiment is the Planck mission, according to the Planck Blue Book \citep{:2006uk}. We use the $100\,\mathrm{GHz}$, $143\,\mathrm{GHz}$, and $217\,\mathrm{GHz}$ channels as science channels. They have a beam of $\theta_\mathrm{fwhm}=9.5'$, $\theta_\mathrm{fwhm}=7.1'$, and $\theta_\mathrm{fwhm}=5'$ and sensitivities of $\sigma_T=2.5\times10^{-6}$, $\sigma_T=2.2\times10^{-6}$, $\sigma_T=4.8\times10^{-6}$ for temperature, and $\sigma_P=4\times10^{-6}$, $\sigma_P=4.2\times10^{-6}$, $\sigma_P=9.8\times10^{-6}$ for polarisation, respectively. We take $f^\mathrm{Planck}_\mathrm{sky}=0.8$ as the sky fraction to account for galactic obstruction, and use a minimum angular mode $\ell_\mathrm{min}=30$ in order to avoid problems with polarisation foregrounds. We discard temperature and polarisation data at $\ell>2000$ to reduce sensitivity to contributions from patchy reionisation and point source contamination \citep[see][and references therein]{Albrecht:2006um}.

\section{Results and Discussion}\label{sec:results}
The set of cosmological parameters we eventually probe is $\vartheta_\alpha=\{\om$, $\ob$, $h$, $\cinf$, $n_s$, $\sigma_8\}$ with fiducial values $\mu_{\vartheta_\alpha}=\{0.28$, $0.045$, $0.7$, $0.0001$, $0.96$, $0.8\}$, respectively. We choose the $\cinf$ value according to previous analyses \citep{Camera:2009uz}. Moreover, it has been shown that this value is slightly favoured by actual data, as obtained in Ref.~\citep{Bertacca:2011in}, where these UDM models have been constrained by cross-correlating CMB measurements with observations from six galaxy catalogues (NVSS, HEAO, 2MASS and SDSS main galaxies, luminous red galaxies, and quasars).

For the matter power spectrum, we use the transfer function of Eisenstein and Hu \citep{Eisenstein:1997ik}. Besides, since no comprehensive study of the non-linear behaviour of perturbations in UDM models have been performed so far, we decide to limit ourselves to the linear case only. Note that this differs from what have been done in Camera et al. \citep{Camera:2010wm}, and our present choice consequently affects the results. However, in the end of this analysis we will also present how a less conservative --- yet reasonable --- approach yields significantly better results. Nonetheless, we prefer to be as conservative as possible. Hence, the physical and angular scales we probe here respectively are $k<k_\mathrm{nl}\simeq0.2\,h\,\mathrm{Mpc}^{-1}$ and $\ell\leq500$. Also note that in the present analysis we are not including the covariance between BAO signal and cosmic shear. The marginal errors $\sigma_{\vartheta_\alpha}$ on the fiducial values $\mu_{\vartheta_\alpha}$ of the cosmological parameters $\vartheta_\alpha$ obtained by following the procedures outlined in Section~\ref{sec:observables} are presented in Table~\ref{tab:errors}. Moreover, we also show the correlation coefficient eq.~\eqref{eq:correlation} of the parameters with $\cinf$, i.e. $r_{\vartheta_\alpha-\cinf}$.
\begin{table}
\hspace{-0.26cm}
\begin{tabular}{|cc|cc|cc|cc|cc|}
\hline
{}&{}&\multicolumn{2}{c|}{\textsc{bao}}&\multicolumn{2}{c|}{\textsc{$\gamma$}}&\multicolumn{2}{c|}{\textsc{bao+$\gamma$}}&\multicolumn{2}{c|}{\textsc{bao+$\gamma$+cmb}}\\
\hline
$\vartheta_\alpha$&$\mu_{\vartheta_\alpha}$ & $\sigma_{\vartheta_\alpha}$ & $r_{\vartheta_\alpha-\cinf}$ & $\sigma_{\vartheta_\alpha}$ & $r_{\vartheta_\alpha-\cinf}$ & $\sigma_{\vartheta_\alpha}$ & $r_{\vartheta_\alpha-\cinf}$ & $\sigma_{\vartheta_\alpha}$ & $r_{\vartheta_\alpha-\cinf}$\\
\hline
$\om$ & $0.28$ & $0.0059$ & $0.49$ & $0.00044$ & $-0.011$ & $0.00043$ & $0.066$ & $0.00037$ & $0.165$ \\
$\ob$ & $0.045$ & $0.0017$ & $0.55$ & $0.0015$ & $0.54$ & $0.00081$ & $0.445$ & $0.00011$ & $-0.079$\\
$h$ & $0.7$ & $0.0037$ & $0.51$ & $0.0043$ & $0.75$ & $0.0021$ & $0.53$ & $0.00055$ & $0.086$\\
$\cinf$ & $0.0001$ & $0.0023$ & $1$ & $0.00023$ & $1$ & $0.00017$ & $1$ & $0.000145$ & $1$ \\
$n_s$ & $0.96$ & $0.010$ & $0.14$ & $0.0078$ & $0.48$ & $0.0048$ & $0.44$ & $0.0015$ & $0.017$\\
$\sigma_8$ & $0.8$ & $-$ & $-$ & $0.0040$ & $0.75$ & $0.00225$ & $0.51$ & $0.00025$ & $-0.15$\\
\hline
\end{tabular}
\caption{Marginal $68\%$ confidence-level errors $\sigma_{\vartheta_\alpha}$ on the fiducial values $\mu_{\vartheta_\alpha}$ of the cosmological parameters $\vartheta_\alpha$, and corresponding correlations $r_{\vartheta_\alpha-\cinf}$ with $\cinf$, for a Euclid-like experiment alone and in combination with Planck.}\label{tab:errors}
\end{table}

Let us start by commenting the BAO survey. This analysis yields good results for the marginal errors on the cosmological parameters. However, it must be noted that the relative marginal error on $\cinf$ is one order of magnitude larger than what obtained by using the shear. This is mainly because the fiducial value of the sound-speed parameter makes the UDM matter power spectrum very similar to what predicted by \lcdm\ --- at least at the linear scales considered in this work. Therefore, the galaxy survey alone is not able of putting tight constraints on $\cinf$, since its effect on the BAO peak of the two-point correlation function \eqref{eq:x^d} is too week to be discriminated against \lcdm. By contrast, the constrain on the dimensionless Hubble constant $h$ is stronger for BAO than for the shear. The reason of that is the sensitivity of the BAO power spectrum to the position of the peak of the two-point correlation function (see Figure~\ref{fig:correlation-function-UDM}).

For what concerns the shear $\gamma$, the errors are thoroughly comparable with the previous case. However, the power-spectrum normalisation $\sigma_8$ is worth a dedicated comment. Indeed, its marginal error is particularly small. The reason is straightforward: weak lensing effects are sourced by the whole large-scale distribution of matter over-densities. This means that it is an estimator of $\om$ directly, without need of any tracer --- like galaxies are. Therefore, it not only can tightly constrain the total matter fraction, but is also independent of the bias. Thus, the shear can lift the well-known degeneracy between $\om$ and $\sigma_8$ which typically plagues galaxy surveys.

The constraints on $\om$ and $\cinf$ coming from BAO+$\gamma$ are mainly dominated by the latter, which is the best. Indeed, its forecast marginal error is $\sim27\%$ smaller than in the $\gamma$-only case. Regarding the other errors, they are diminished almost by a factor two, which is a very good result. Besides, the marginalisation process also tightens the shear constraint on $\sigma_8$, albeit the power-spectrum normalisation has been marginalised over in the BAO Fisher matrix.

The best marginal errors are obtained when we combine both Euclid probes together with the priors from Planck. For the latter, we calculate the corresponding Fisher matrix $\mathbf F^\mathrm{CMB}$ \eqref{eqn:cmbfisher} by using \textsc{camb} \citep{Lewis:1999bs}.\footnote{http://camb.info/} Since the class of UDM models we scrutinise here is explicitly constructed to have no tension with CMB measurements, we cannot use Planck to directly probe the scalar-field speed of sound. Nevertheless, its constraining potential on the other cosmological parameters is far well known. Furthermore, as can be seen in the rightmost columns of Table~\ref{tab:errors}, Planck priors on the \lcdm\ parameters yield an enhancement on the $\cinf$ error anyway. This is because $\sigma_{\vartheta_\alpha}$'s are the $1\sigma$ confidence level marginalised over all the other parameters.

\begin{figure}
\centering
\includegraphics[width=0.45\textwidth]{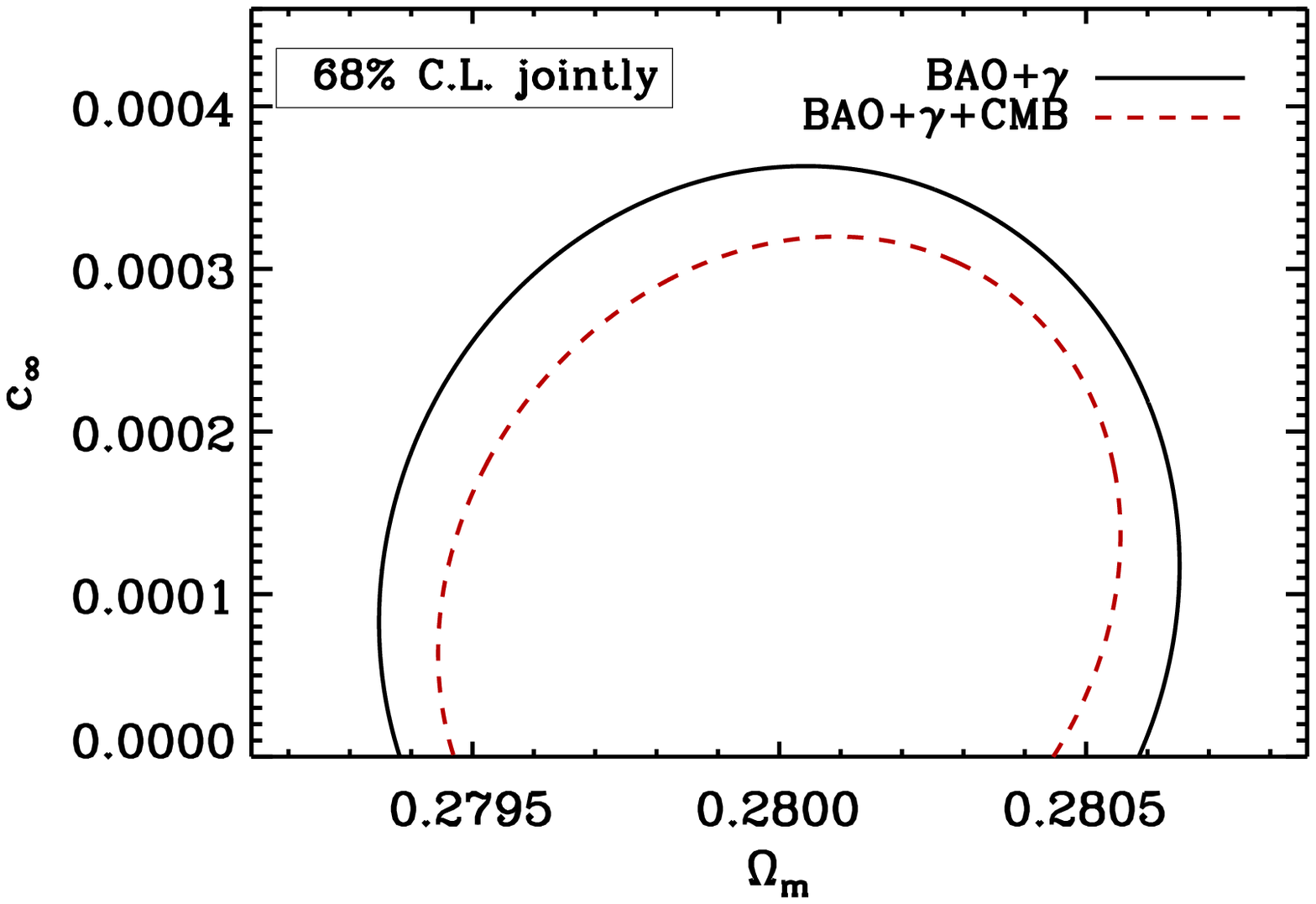}\\
\includegraphics[width=0.45\textwidth]{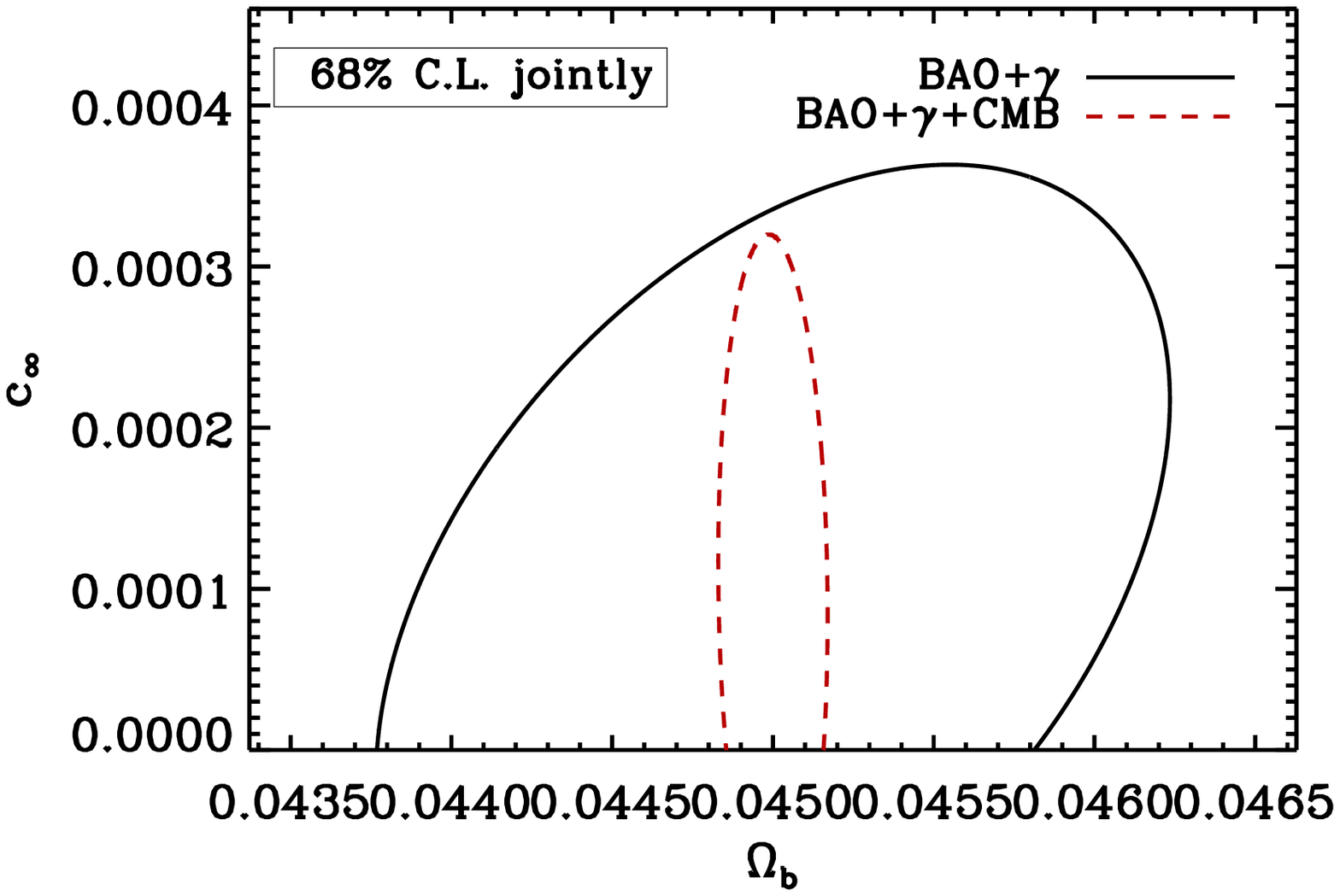}
\includegraphics[width=0.45\textwidth]{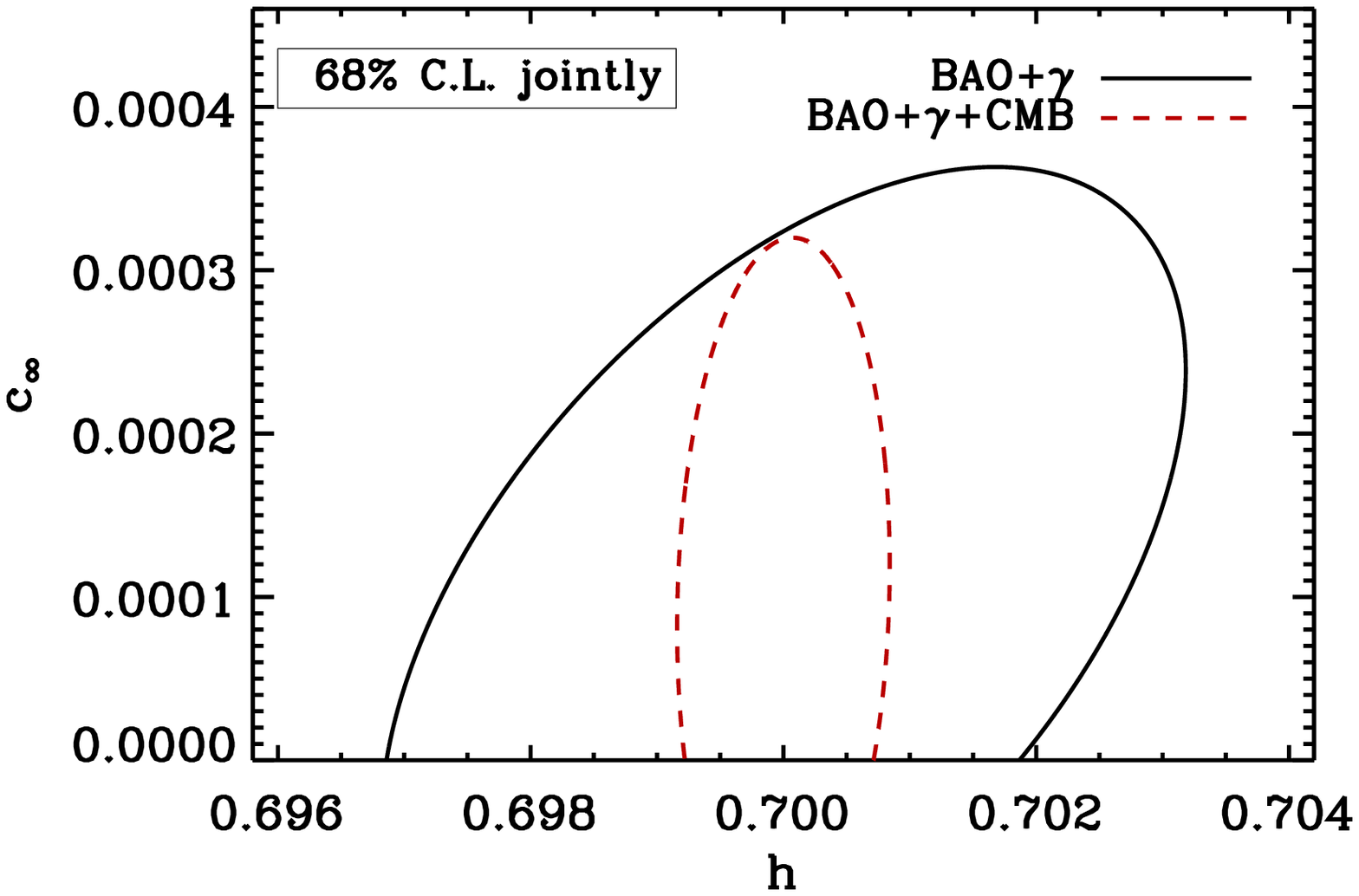}\\
\includegraphics[width=0.45\textwidth]{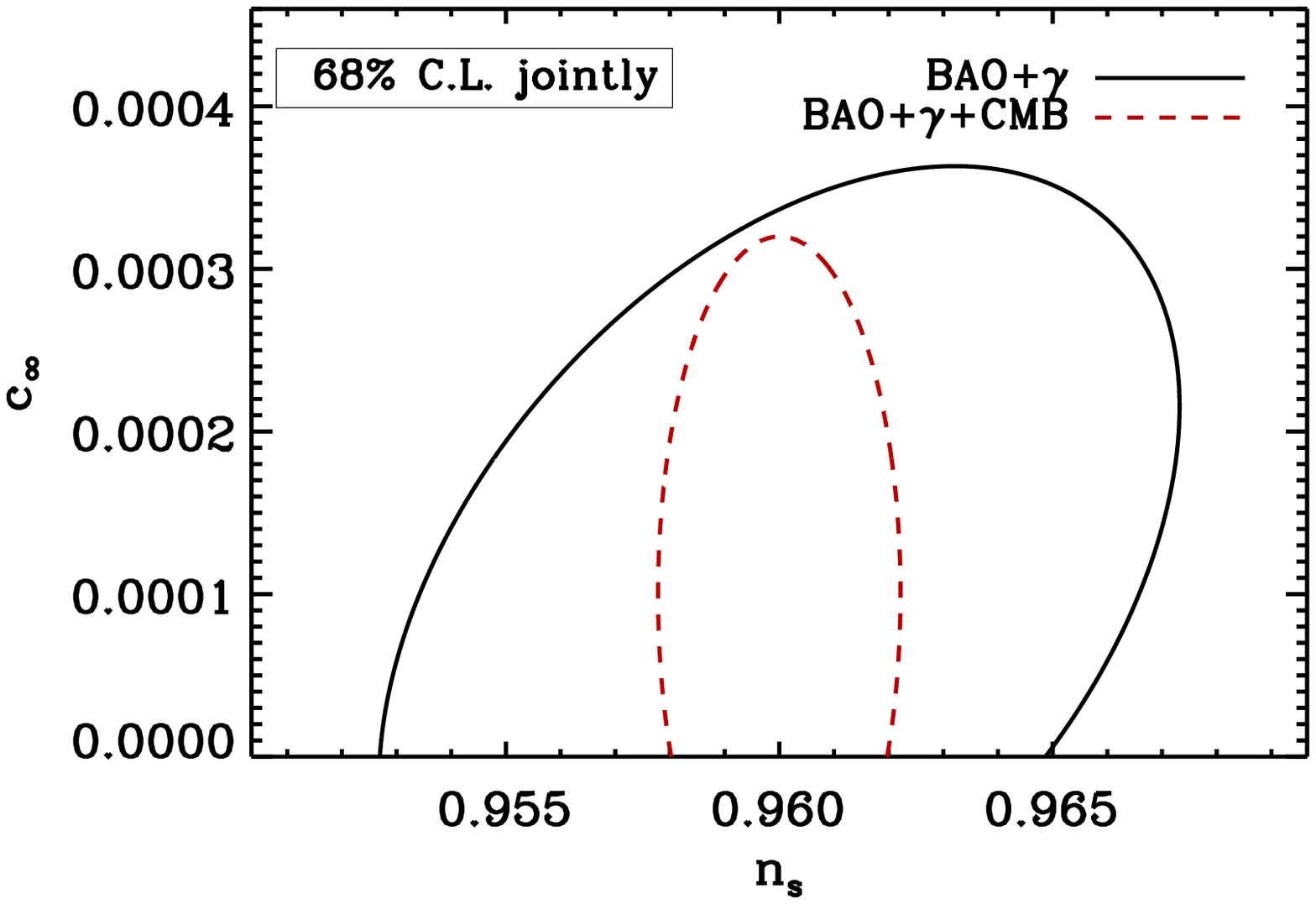}
\includegraphics[width=0.45\textwidth]{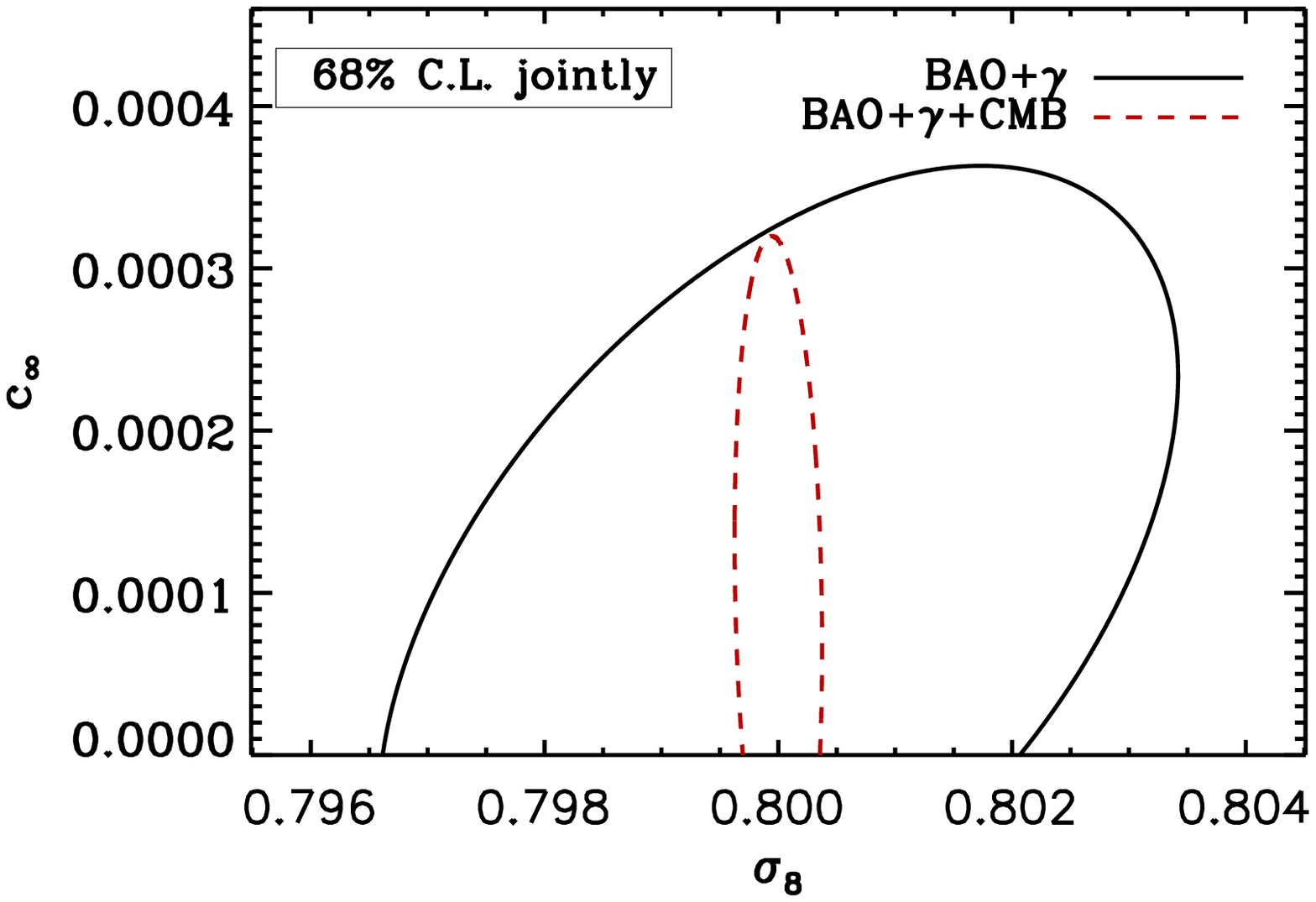}
\caption{Two-parameter projected $68\%$ contours in the $(c_\infty,\,\vartheta_{\alpha'})$-plane, with $\vartheta_{\alpha'}=\{\Omega_m$, $\Omega_b$, $h,n_s,\sigma_8\}$. The solid black line corresponds to BAO+$\gamma$ data from an Euclid-like survey. The dashed red line is obtained with the further addition of Planck priors.}
\label{contour1}
\end{figure}

To better understand these results on $\cinf$, we show in Figure~\ref{contour1} the $68\%$ two-parameter contours in the $(c_\infty,\,\vartheta_{\alpha'})$-plane, with $\vartheta_{\alpha'}=\{\Omega_m$, $\Omega_b$, $h,n_s,\sigma_8\}$. The solid black line corresponds to BAO+$\gamma$ data from an Euclid-like survey. The dashed red line is obtained with the further addition of Planck priors.

Looking at the correlations $r_{\vartheta_\alpha-\cinf}$, the sound speed appears degenerate with respect to some of the other parameters, when we consider the shear only. Particularly, this degeneracy shows up with the slope $n_s$ of the primordial power spectrum and its normalisation $\sigma_8$. Actually, this fact was expected; indeed, Camera et al. \citep{Camera:2009uz} have demonstrated that, if we restrict ourselves to the linear r\'egime of perturbations --- namely, for small $\ell$'s ---, the suppression of growth due to a large value of $\cinf$ can be healed by tuning $n_s$ and $\sigma_8$. As a result, their correlations with $\cinf$ are larger. Contrarily, this does not happen in the BAO case, since here we marginalise over the power-spectrum normalisation. Finally, when we add the Planck priors, almost all the correlations drop down, since the use of CMB data here is exactly for lifting the degeneracies between parameters. This is clearly visible in Figure~\ref{contour1}.

We want to emphasise that these results have been obtained by using only linear (physical or angular) scales. This means that they are completely reliable and robust, since in that r\'egime the growth of perturbation for the dark-matter-like component of the UDM scalar field is well understood. Moreover, it is also worth underlining once more that the $\cinf$ fiducial value we use is in agreement with all the relevant cosmological data which \lcdm\ has been fitted against. Such value is also somehow favoured by several of the presently-available galaxy catalogues \citep{Bertacca:2011in}. Unfortunately, the price of this agreement is that this UDM model is hardly distinguishable from pure \lcdm, as it is demonstrated by our all-inclusive forecast accuracy $\sigma_\cinf=0.0001$. Hence, the development of a non-linear framework for UDM models is one the major priorities for allowing us to better test and either confirm or rule out this class of alternative cosmologies.

As a final remark, we show how our approach would yield much better results if we included some amount of non-linear information. To do this, we follow Camera et al. \citep{Camera:2010wm}, where the authors have assumed that a \textsc{halofit} non-linear power spectrum may well behave for small-$\cinf$ UDM models. Although this assumption is arbitrary and not supported by any theoretical or numerical piece of evidence, it is a fairly reasonable approximation. Indeed, the power spectrum of the clustering component of the scalar field in a UDM model with $\cinf=0.0001$ differs from the standard \lcdm\ matter power spectrum by less than $10\%$ still at mildly non-linear scales such as $k=0.3\,h\,\mathrm{Mpc}^{-1}$. Table~\ref{tab:errors_nl} shows the $1\sigma$ parameter errors obtained by performing the cosmic-shear Fisher analysis up to $\ell_\mathrm{max}=1000$ or $5000$, in comparison with the linear result with $\ell_\mathrm{max}=500$. By doing so, we immediately are able to detect significant departures from the \lcdm\ expected signal and thus lift such a model degeneracy. Indeed, we find $\cinf=0.0001\pm0.00007$, when we include non-linear angular scales up to $\ell_\mathrm{max}=1000$, or even $\cinf=0.0001\pm3\cdot10^{-6}$, when $\ell_\mathrm{max}=5000$. This is particularly important because Euclid will cover those non-linear scales.
\begin{table}
\centering
\begin{tabular}{|cc|c|c|c|}
\hline
{}&{}& $\ell_\mathrm{max}=500$ & $\ell_\mathrm{max}=1000$ & $\ell_\mathrm{max}=5000$ \\
\hline
$\vartheta_\alpha$&$\mu_{\vartheta_\alpha}$ & $\sigma_{\vartheta_\alpha}$ & $\sigma_{\vartheta_\alpha}$ & $\sigma_{\vartheta_\alpha}$ \\
\hline
$\om$ & $0.28$ & $0.0004$ & $0.0003$ & $0.00007$ \\
$\ob$ & $0.045$ & $0.001$ & $0.0007$ & $0.0004$ \\
$h$ & $0.7$ & $0.004$ & $0.002$ & $0.0003$ \\
$\cinf$ & $0.0001$ & $0.0002$ & $0.00007$ & $3\cdot10^{-6}$\\
$n_s$ & $0.96$ & $0.008$ & $0.003$ & $0.001$ \\
$\sigma_8$ & $0.8$ & $0.004$ & $0.001$ & $0.0002$ \\
\hline
\end{tabular}
\caption{Marginal $1\sigma$ errors $\sigma_{\vartheta_\alpha}$ on the fiducial values $\mu_{\vartheta_\alpha}$ of the cosmological parameters $\vartheta_\alpha$ for cosmic-shear Euclid-like experiment with the inclusion of non-linear angular scales up to $\ell_\mathrm{max}=1000$ (fourth column) or $5000$ (fifth column), in comparison with the linear result with $\ell_\mathrm{max}=500$ (third column).}\label{tab:errors_nl}
\end{table}

\section{Conclusions}\label{sec:conclusions}
In this paper, we scrutinise a class of cosmological models where --- conversely to \lcdm\ --- dark matter and dark energy are two aspects of the same ``dark fluid.'' Specifically, an exotic scalar field presents an energy density containing both a dark-matter-like term and a constant term which could play the r\^ole of the cosmological constant $\Lambda$ \citep[see][for an exhaustive review]{Bertacca:2010ct}. We refer to such cosmologies as Unified Dark Matter (UDM), and we focus on a cosmologically-viable family of UDM models \citep{Bertacca:2008uf}, where the only extra-\lcdm\ parameter is $\cinf$, viz. the value of the scalar-field sound speed when $a\to\infty$. We use a fiducial value which has been obtained by Bertacca et al. \citep{Bertacca:2011in} by cross-correlating CMB measurements with observations from six galaxy catalogues.

The main features of these UDM models is indeed the presence of such non-negligible speed of sound for the perturbations of the dark-matter component of the scalar field. This implies an effective Jeans length for the Newtonian potential, which causes a strong time and scale dependence on the growth of dark-matter-like fluctuations. Particularly, perturbations which form on scales smaller than the inverse of the Jeans length cannot cluster. Furthermore, the scalar-field sound speed is a time-dependent quantity, it therefore introduces an additional redshift dependence in the growth of cosmic structures.

Given these peculiarities, we decide to exploit the potentialities of the future generation of wide-field, large-scale surveys to constrain these UDM models. A similar approach has been also adopted by Xu, Wang and Noh \citep{Xu:2011bp}, whose UDM model has instead a constant, adiabatic speed of sound \citep[see also][]{Aviles:2011ak}. Here, we focus on the Euclid mission \citep{EditorialTeam:2011mu}. Indeed, it will be a powerful tool, as it involves both a spectroscopic galaxy survey and an imaging photometric survey. We make use of the former to study the BAO signatures expected in UDM models, whilst the latter is utilised for analysing the cosmic shear $\gamma$, i.e. the weak-lensing signal due to the potential wells of the large-scale structure of the Universe. These probes perfectly suit the pursue; BAO are caused by oscillations in the baryon-photon fluid before recombination and, as a matter of fact, the clustering suppression of the scalar-field perturbations below the Jeans length acts by making the Newtonian potential strongly oscillating as well. On the other hand, the  cosmic shear encodes the large-scale distribution of under- and over-densities in the cosmos, which can be substantially different in UDM models compared to \lcdm. Finally, we also take advantage of the ongoing Planck mission, which will yield extremely accurate data on both the temperature anisotropies of the CMB and its polarisation.

We decide to not include non-linear scales in our analysis. Indeed, in UDM models there is, so far, no theory available able of predicting the growth of perturbations in the non-linear r\'egime. Therefore, we consider only $k<k_\mathrm{nl}\simeq0.2\,h\,\mathrm{Mpc}^{-1}$ and $\ell\leq\ell_\mathrm{max}=500$. This strongly affects our results. The major problem is that a UDM model with our fiducial $\cinf=0.0001$ is hardly distinguishable from a standard \lcdm\ model. Though this is good because there is thus no tension with the currently present data, it makes it very hard to discriminate amongst the models. However, we still find that UDM models can be constrained to a certain degree of accuracy by the joint venture of Eulid and Planck, with an eventual relative marginal error of order $1$. This is a remarkably good result, given the UDM power spectrum almost degenerate with \lcdm, on linear scales.

As a ultimate result, we explore what could be done if some amount of non-linear information were included in the analysis. According to Camera et al. \citep{Camera:2010wm} we assume that \textsc{halofit} fitting formul\ae\ well reproduce the UDM non-linear behaviour. This is certainly an arbitrary assumption, but it is also reasonable, since the power spectrum of the clustering component of the scalar field in a UDM model with $\cinf=0.0001$ differs from the standard \lcdm\ matter power spectrum by less than $10\%$ still at mildly non-linear scales. By doing so, we found extremely better results, with marginal errors on $\cinf$ of order $0.7$ or even $0.03$.

\acknowledgments SC acknowledges support from FCT-Portugal under grant PTDC/FIS/100170/2008. SC want also to thank the Dipartimento di Astronomia, Universit\`a di Bologna for hospitality and financial support during the development of this project. CC and LM acknowledge financial contributions from contracts ASI-INAF I/023/05/0, ASI-INAF I/088/06/0, ASI I/016/07/0 `COFIS,' ASI `Euclid-DUNE' I/064/08/0, ASI-Uni Bologna-Astronomy Dept. `Euclid-NIS' I/039/10/0, and PRIN MIUR ``Dark energy and cosmology with large galaxy surveys.''

\bibliographystyle{JHEP}
\bibliography{/home/stefano/Documents/LaTeX/Bibliography}

\providecommand{\href}[2]{#2}\begingroup\raggedright\begin{thebibliography}{10}

\bibitem{deBernardis:2000gy}
{\bf Boomerang} Collaboration, P.~de~Bernardis {\em et.~al.}, {\it {A Flat
  Universe from High-Resolution Maps of the Cosmic Microwave Background
  Radiation}},  {\em Nature} {\bf 404} (2000) 955--959,
  [\href{http://xxx.lanl.gov/abs/astro-ph/0004404}{{\tt astro-ph/0004404}}].

\bibitem{Stompor:2001xf}
R.~Stompor {\em et.~al.}, {\it {Cosmological implications of the MAXIMA-I high
  resolution Cosmic Microwave Background anisotropy measurement}},  {\em
  Astrophys. J.} {\bf 561} (2001) L7--L10,
  [\href{http://xxx.lanl.gov/abs/astro-ph/0105062}{{\tt astro-ph/0105062}}].

\bibitem{Netterfield:2001yq}
{\bf Boomerang} Collaboration, C.~B. Netterfield {\em et.~al.}, {\it {A
  measurement by BOOMERANG of multiple peaks in the angular power spectrum of
  the cosmic microwave background}},  {\em Astrophys. J.} {\bf 571} (2002)
  604--614, [\href{http://xxx.lanl.gov/abs/astro-ph/0104460}{{\tt
  astro-ph/0104460}}].

\bibitem{Rebolo:2004vp}
R.~Rebolo {\em et.~al.}, {\it {Cosmological parameter estimation using Very
  Small Array data out to l=1500}},  {\em Mon. Not. Roy. Astron. Soc.} {\bf
  353} (2004) 747--759, [\href{http://xxx.lanl.gov/abs/astro-ph/0402466}{{\tt
  astro-ph/0402466}}].

\bibitem{Bennett:2003bz}
{\bf WMAP} Collaboration, C.~L. Bennett {\em et.~al.}, {\it {First Year
  Wilkinson Microwave Anisotropy Probe (WMAP) Observations: Preliminary Maps
  and Basic Results}},  {\em Astrophys. J. Suppl.} {\bf 148} (2003) 1,
  [\href{http://xxx.lanl.gov/abs/astro-ph/0302207}{{\tt astro-ph/0302207}}].

\bibitem{Spergel:2003cb}
{\bf WMAP} Collaboration, D.~N. Spergel {\em et.~al.}, {\it {First Year
  Wilkinson Microwave Anisotropy Probe (WMAP) Observations: Determination of
  Cosmological Parameters}},  {\em Astrophys. J. Suppl.} {\bf 148} (2003) 175,
  [\href{http://xxx.lanl.gov/abs/astro-ph/0302209}{{\tt astro-ph/0302209}}].

\bibitem{Komatsu:2008hk}
{\bf WMAP} Collaboration, E.~Komatsu {\em et.~al.}, {\it {Five-Year Wilkinson
  Microwave Anisotropy Probe (WMAP) Observations:Cosmological Interpretation}},
   {\em Astrophys. J. Suppl.} {\bf 180} (2009) 330--376,
  [\href{http://xxx.lanl.gov/abs/0803.0547}{{\tt arXiv:0803.0547}}].

\bibitem{Perlmutter:1996ds}
{\bf Supernova Cosmology Project} Collaboration, S.~Perlmutter {\em et.~al.},
  {\it {Measurements of the Cosmological Parameters Omega and Lambda from the
  First 7 Supernovae at z $\geq$ 0.35}},  {\em Astrophys. J.} {\bf 483} (1997)
  565, [\href{http://xxx.lanl.gov/abs/astro-ph/9608192}{{\tt
  astro-ph/9608192}}].

\bibitem{Riess:1998cb}
{\bf Supernova Search Team} Collaboration, A.~G. Riess {\em et.~al.}, {\it
  {Observational Evidence from Supernovae for an Accelerating Universe and a
  Cosmological Constant}},  {\em Astron. J.} {\bf 116} (1998) 1009--1038,
  [\href{http://xxx.lanl.gov/abs/astro-ph/9805201}{{\tt astro-ph/9805201}}].

\bibitem{Schmidt:1998ys}
{\bf Supernova Search Team} Collaboration, B.~P. Schmidt {\em et.~al.}, {\it
  {The High-Z Supernova Search: Measuring Cosmic Deceleration and Global Cur
  vature of the Universe Using Type Ia Supernovae}},  {\em Astrophys. J.} {\bf
  507} (1998) 46--63, [\href{http://xxx.lanl.gov/abs/astro-ph/9805200}{{\tt
  astro-ph/9805200}}].

\bibitem{Perlmutter:1998np}
{\bf Supernova Cosmology Project} Collaboration, S.~Perlmutter {\em et.~al.},
  {\it {Measurements of Omega and Lambda from 42 High-Redshift Supernovae}},
  {\em Astrophys. J.} {\bf 517} (1999) 565--586,
  [\href{http://xxx.lanl.gov/abs/astro-ph/9812133}{{\tt astro-ph/9812133}}].

\bibitem{Knop:2003iy}
{\bf Supernova Cosmology Project} Collaboration, R.~A. Knop {\em et.~al.}, {\it
  {New Constraints on $\Omega_M$, $\Omega_\Lambda$, and w from an Independent
  Set of Eleven High-Redshift Supernovae Observed with HST}},  {\em Astrophys.
  J.} {\bf 598} (2003) 102,
  [\href{http://xxx.lanl.gov/abs/astro-ph/0309368}{{\tt astro-ph/0309368}}].

\bibitem{Tonry:2003zg}
{\bf Supernova Search Team} Collaboration, J.~L. Tonry {\em et.~al.}, {\it
  {Cosmological Results from High-z Supernovae}},  {\em Astrophys. J.} {\bf
  594} (2003) 1--24, [\href{http://xxx.lanl.gov/abs/astro-ph/0305008}{{\tt
  astro-ph/0305008}}].

\bibitem{Riess:2004n}
{\bf Supernova Search Team} Collaboration, A.~G. Riess {\em et.~al.}, {\it
  {Type Ia Supernova Discoveries at z>1 From the Hubble Space Telescope:
  Evidence for Past Deceleration and Constraints on Dark Energy Evolution}},
  {\em Astrophys. J.} {\bf 607} (2004) 665--687,
  [\href{http://xxx.lanl.gov/abs/astro-ph/0402512}{{\tt astro-ph/0402512}}].

\bibitem{Astier:2005qq}
{\bf The SNLS} Collaboration, P.~Astier {\em et.~al.}, {\it {The Supernova
  Legacy Survey: Measurement of Omega\_M, Omega\_Lambda and w from the First
  Year Data Set}},  {\em Astron. Astrophys.} {\bf 447} (2006) 31--48,
  [\href{http://xxx.lanl.gov/abs/astro-ph/0510447}{{\tt astro-ph/0510447}}].

\bibitem{WoodVasey:2007jb}
{\bf ESSENCE} Collaboration, W.~M. Wood-Vasey {\em et.~al.}, {\it
  {Observational Constraints on the Nature of the Dark Energy: First
  Cosmological Results from the ESSENCE Supernova Survey}},  {\em Astrophys.
  J.} {\bf 666} (2007) 694--715,
  [\href{http://xxx.lanl.gov/abs/astro-ph/0701041}{{\tt astro-ph/0701041}}].

\bibitem{Percival:2002gq}
{\bf The 2dFGRS Team} Collaboration, W.~J. Percival {\em et.~al.}, {\it
  {Parameter constraints for flat cosmologies from CMB and 2dFGRS power
  spectra}},  {\em Mon. Not. Roy. Astron. Soc.} {\bf 337} (2002) 1068,
  [\href{http://xxx.lanl.gov/abs/astro-ph/0206256}{{\tt astro-ph/0206256}}].

\bibitem{Pope:2004cc}
{\bf The SDSS} Collaboration, A.~C. Pope {\em et.~al.}, {\it {Cosmological
  Parameters from Eigenmode Analysis of Sloan Digital Sky Survey Galaxy
  Redshifts}},  {\em Astrophys. J.} {\bf 607} (2004) 655--660,
  [\href{http://xxx.lanl.gov/abs/astro-ph/0401249}{{\tt astro-ph/0401249}}].

\bibitem{Zwicky:1933gu}
F.~Zwicky, {\it {Spectral displacement of extra galactic nebulae}},  {\em Helv.
  Phys. Acta} {\bf 6} (1933) 110--127.

\bibitem{Zwicky:1937zza}
F.~Zwicky, {\it {On the Masses of Nebulae and of Clusters of Nebulae}},  {\em
  Astrophys. J.} {\bf 86} (1937) 217--246.

\bibitem{Dodelson:2001ux}
{\bf SDSS} Collaboration, S.~Dodelson {\em et.~al.}, {\it {The
  three-dimensional power spectrum from angular clustering of galaxies in early
  SDSS data}},  {\em Astrophys. J.} {\bf 572} (2001) 140--156,
  [\href{http://xxx.lanl.gov/abs/astro-ph/0107421}{{\tt astro-ph/0107421}}].

\bibitem{Hawkins:2002sg}
E.~Hawkins {\em et.~al.}, {\it {The 2dF Galaxy Redshift Survey: correlation
  functions, peculiar velocities and the matter density of the Universe}},
  {\em Mon. Not. Roy. Astron. Soc.} {\bf 346} (2003) 78,
  [\href{http://xxx.lanl.gov/abs/astro-ph/0212375}{{\tt astro-ph/0212375}}].

\bibitem{Spergel:2006hy}
{\bf WMAP} Collaboration, D.~N. Spergel {\em et.~al.}, {\it {Wilkinson
  Microwave Anisotropy Probe (WMAP) three year results: Implications for
  cosmology}},  {\em Astrophys. J. Suppl.} {\bf 170} (2007) 377,
  [\href{http://xxx.lanl.gov/abs/astro-ph/0603449}{{\tt astro-ph/0603449}}].

\bibitem{Riess:2006fw}
A.~G. Riess {\em et.~al.}, {\it {New Hubble Space Telescope Discoveries of Type
  Ia Supernovae at $z > 1$: Narrowing Constraints on the Early Behavior of Dark
  Energy}},  {\em Astrophys. J.} {\bf 659} (2007) 98--121,
  [\href{http://xxx.lanl.gov/abs/astro-ph/0611572}{{\tt astro-ph/0611572}}].

\bibitem{Tisserand:2006zx}
{\bf EROS-2} Collaboration, P.~Tisserand {\em et.~al.}, {\it {Limits on the
  Macho Content of the Galactic Halo from the EROS-2 Survey of the Magellanic
  Clouds}},  {\em Astron. Astrophys.} {\bf 469} (2007) 387--404,
  [\href{http://xxx.lanl.gov/abs/astro-ph/0607207}{{\tt astro-ph/0607207}}].

\bibitem{Wyrzykowski:2009ep}
L.~Wyrzykowski, S.~Kozlowski, J.~Skowron, V.~Belokurov, M.~Smith, {\em
  et.~al.}, {\it {The OGLE View of Microlensing towards the Magellanic Clouds.
  I. A Trickle of Events in the OGLE-II LMC data}},  {\em Mon. Not. Roy.
  Astron. Soc.} {\bf 397} (2009) 1228--1242,
  [\href{http://xxx.lanl.gov/abs/0905.2044}{{\tt arXiv:0905.2044}}].

\bibitem{Bertone:2004pz}
G.~Bertone, D.~Hooper, and J.~Silk, {\it {Particle dark matter: Evidence,
  candidates and constraints}},  {\em Phys. Rept.} {\bf 405} (2005) 279--390,
  [\href{http://xxx.lanl.gov/abs/hep-ph/0404175}{{\tt hep-ph/0404175}}].

\bibitem{Fornengo:2006yy}
N.~Fornengo, {\it {Status and perspectives of indirect and direct dark matter
  searches}},  {\em Adv. Space Res.} {\bf 41} (2008) 2010--2018,
  [\href{http://xxx.lanl.gov/abs/astro-ph/0612786}{{\tt astro-ph/0612786}}].

\bibitem{Feng:2010gw}
J.~L. Feng, {\it {Dark Matter Candidates from Particle Physics and Methods of
  Detection}},  {\em Ann. Rev. Astron. Astrophys.} {\bf 48} (2010) 495,
  [\href{http://xxx.lanl.gov/abs/1003.0904}{{\tt arXiv:1003.0904}}].

\bibitem{Weinberg:1988cp}
S.~Weinberg, {\it {The Cosmological Constant Problem}},  {\em Rev.Mod.Phys.}
  {\bf 61} (1989) 1--23. Morris Loeb Lectures in Physics, Harvard University,
  May 2, 3, 5, and 10, 1988.

\bibitem{Carroll:2000fy}
S.~M. Carroll, {\it {The Cosmological Constant}},  {\em Living Rev. Rel.} {\bf
  4} (2001) 1, [\href{http://xxx.lanl.gov/abs/astro-ph/0004075}{{\tt
  astro-ph/0004075}}].

\bibitem{Bertacca:2010ct}
D.~Bertacca, N.~Bartolo, and S.~Matarrese, {\it {Unified Dark Matter Scalar
  Field Models}},  {\em Adv. Astron.} {\bf 2010} (2010) 904379,
  [\href{http://xxx.lanl.gov/abs/1008.0614}{{\tt arXiv:1008.0614}}]. *
  Temporary entry *.

\bibitem{Kamenshchik:2001cp}
A.~Y. Kamenshchik, U.~Moschella, and V.~Pasquier, {\it {An alternative to
  quintessence}},  {\em Phys. Lett.} {\bf B511} (2001) 265--268,
  [\href{http://xxx.lanl.gov/abs/gr-qc/0103004}{{\tt gr-qc/0103004}}].

\bibitem{Bilic:2001cg}
N.~Bilic, G.~B. Tupper, and R.~D. Viollier, {\it {Unification of dark matter
  and dark energy: The inhomogeneous Chaplygin gas}},  {\em Phys. Lett.} {\bf
  B535} (2002) 17--21, [\href{http://xxx.lanl.gov/abs/astro-ph/0111325}{{\tt
  astro-ph/0111325}}].

\bibitem{Bento:2002ps}
M.~C. Bento, O.~Bertolami, and A.~A. Sen, {\it {Generalized Chaplygin gas,
  accelerated expansion and dark energy-matter unification}},  {\em Phys. Rev.}
  {\bf D66} (2002) 043507, [\href{http://xxx.lanl.gov/abs/gr-qc/0202064}{{\tt
  gr-qc/0202064}}].

\bibitem{Balbi:2007mz}
A.~Balbi, M.~Bruni, and C.~Quercellini, {\it {Lambda-alpha DM: Observational
  constraints on unified dark matter with constant speed of sound}},  {\em
  Phys. Rev.} {\bf D76} (2007) 103519,
  [\href{http://xxx.lanl.gov/abs/astro-ph/0702423}{{\tt astro-ph/0702423}}].

\bibitem{Quercellini:2007ht}
C.~Quercellini, M.~Bruni, and A.~Balbi, {\it {Affine equation of state from
  quintessence and k-essence fields}},  {\em Class. Quant. Grav.} {\bf 24}
  (2007) 5413--5426, [\href{http://xxx.lanl.gov/abs/0706.3667}{{\tt
  arXiv:0706.3667}}].

\bibitem{Scherrer:2004au}
R.~J. Scherrer, {\it {Purely kinetic k-essence as unified dark matter}},  {\em
  Phys. Rev. Lett.} {\bf 93} (2004) 011301,
  [\href{http://xxx.lanl.gov/abs/astro-ph/0402316}{{\tt astro-ph/0402316}}].

\bibitem{Bertacca:2007ux}
D.~Bertacca, S.~Matarrese, and M.~Pietroni, {\it {Unified dark matter in scalar
  field cosmologies}},  {\em Mod. Phys. Lett.} {\bf A22} (2007) 2893--2907,
  [\href{http://xxx.lanl.gov/abs/astro-ph/0703259}{{\tt astro-ph/0703259}}].

\bibitem{Chimento:2009nj}
L.~P. Chimento, R.~Lazkoz, and I.~Sendra, {\it {DBI models for the unification
  of dark matter and dark energy}},  {\em arXiv:0904.1114} (2009)
  [\href{http://xxx.lanl.gov/abs/0904.1114}{{\tt arXiv:0904.1114}}].

\bibitem{Piattella:2009kt}
O.~F. Piattella, D.~Bertacca, M.~Bruni, and D.~Pietrobon, {\it {Unified Dark
  Matter models with fast transition}},  {\em JCAP} {\bf 1001} (2010) 014,
  [\href{http://xxx.lanl.gov/abs/0911.2664}{{\tt arXiv:0911.2664}}].

\bibitem{Bertacca:2010mt}
D.~Bertacca, M.~Bruni, O.~F. Piattella, and D.~Pietrobon, {\it {Unified Dark
  Matter scalar field models with fast transition}},  {\em JCAP} {\bf 1102}
  (2011) 018, [\href{http://xxx.lanl.gov/abs/1011.6669}{{\tt
  arXiv:1011.6669}}].

\bibitem{Bertacca:2008uf}
D.~Bertacca, N.~Bartolo, A.~Diaferio, and S.~Matarrese, {\it {How the Scalar
  Field of Unified Dark Matter Models Can Cluster}},  {\em JCAP} {\bf 0810}
  (2008) 023, [\href{http://xxx.lanl.gov/abs/0807.1020}{{\tt
  arXiv:0807.1020}}].

\bibitem{Bertacca:2007cv}
D.~Bertacca and N.~Bartolo, {\it {ISW effect in Unified Dark Matter Scalar
  Field Cosmologies: an analytical approach}},  {\em JCAP} {\bf 0711} (2007)
  026, [\href{http://xxx.lanl.gov/abs/0707.4247}{{\tt arXiv:0707.4247}}].

\bibitem{Camera:2009uz}
S.~Camera, D.~Bertacca, A.~Diaferio, N.~Bartolo, and S.~Matarrese, {\it {Weak
  lensing signal in unified dark matter models}},  {\em Mon. Not. Roy. Astron.
  Soc.} {\bf 399} (2009) 1995--2003,
  [\href{http://xxx.lanl.gov/abs/0902.4204}{{\tt arXiv:0902.4204}}].

\bibitem{Camera:2010wm}
S.~Camera, T.~D. Kitching, A.~F. Heavens, D.~Bertacca, and A.~Diaferio, {\it
  {Measuring Unified Dark Matter with 3D cosmic shear}},  {\em Mon. Not. Roy.
  Astron. Soc.} {\bf 415} (2011) 399--409,
  [\href{http://xxx.lanl.gov/abs/1002.4740}{{\tt arXiv:1002.4740}}].

\bibitem{Bertacca:2011in}
D.~Bertacca, A.~Raccanelli, O.~F. Piattella, D.~Pietrobon, N.~Bartolo, {\em
  et.~al.}, {\it {CMB-Galaxy correlation in Unified Dark Matter Scalar Field
  Cosmologies}},  {\em JCAP} {\bf 1103} (2011) 039,
  [\href{http://xxx.lanl.gov/abs/1102.0284}{{\tt arXiv:1102.0284}}].

\bibitem{EditorialTeam:2011mu}
{\bf Euclid} Collaboration, R.~Laureijs, J.~Amiaux, S.~Arduini, J.-L. Augueres,
  {\em et.~al.}, {\it {Euclid Definition Study Report}},  {\em ESA-SRE} {\bf
  12} (2011) [\href{http://xxx.lanl.gov/abs/1110.3193}{{\tt arXiv:1110.3193}}].

\bibitem{:2006uk}
{\bf Planck} Collaboration, {\it {The Scientific programme of Planck}},  {\em
  ESA-SCI} {\bf 1} (2005) [\href{http://xxx.lanl.gov/abs/astro-ph/0604069}{{\tt
  astro-ph/0604069}}].

\bibitem{Born:1934gh}
M.~Born and L.~Infeld, {\it {Foundations of the new field theory}},  {\em Proc.
  Roy. Soc. Lond.} {\bf A144} (1934) 425--451.

\bibitem{Padmanabhan:2002sh}
T.~Padmanabhan and T.~Choudhury, {\it {Can the clustered dark matter and the
  smooth dark energy arise from the same scalar field?}},  {\em Phys. Rev.}
  {\bf D66} (2002) 081301, [\href{http://xxx.lanl.gov/abs/hep-th/0205055}{{\tt
  hep-th/0205055}}].

\bibitem{Abramo:2003cp}
L.~R.~W. Abramo and F.~Finelli, {\it {Cosmological dynamics of the tachyon with
  an inverse power-law potential}},  {\em Phys. Lett.} {\bf B575} (2003)
  165--171, [\href{http://xxx.lanl.gov/abs/astro-ph/0307208}{{\tt
  astro-ph/0307208}}].

\bibitem{Abramo:2004ji}
L.~R. Abramo, F.~Finelli, and T.~S. Pereira, {\it {Constraining Born-Infeld
  models of dark energy with CMB anisotropies}},  {\em Phys. Rev.} {\bf D70}
  (2004) 063517, [\href{http://xxx.lanl.gov/abs/astro-ph/0405041}{{\tt
  astro-ph/0405041}}].

\bibitem{Sen:2002nu}
A.~Sen, {\it {Rolling Tachyon}},  {\em JHEP} {\bf 04} (2002) 048,
  [\href{http://xxx.lanl.gov/abs/hep-th/0203211}{{\tt hep-th/0203211}}].

\bibitem{Sen:2002in}
A.~Sen, {\it {Tachyon matter}},  {\em JHEP} {\bf 07} (2002) 065,
  [\href{http://xxx.lanl.gov/abs/hep-th/0203265}{{\tt hep-th/0203265}}].

\bibitem{Sen:2002vv}
A.~Sen, {\it {Time evolution in open string theory}},  {\em JHEP} {\bf 10}
  (2002) 003, [\href{http://xxx.lanl.gov/abs/hep-th/0207105}{{\tt
  hep-th/0207105}}].

\bibitem{Bardeen:1980kt}
J.~M. Bardeen, {\it {Gauge Invariant Cosmological Perturbations}},  {\em Phys.
  Rev.} {\bf D22} (1980) 1882--1905.

\bibitem{Mukhanov:2005sc}
V.~Mukhanov, {\em Physical foundations of cosmology}.
\newblock 2005.
\newblock ~Cambridge, UK: Univ. Pr. (2005) 421 p.

\bibitem{Fisher:1935}
R.~A. Fisher {\em J. Roy. Stat. Soc.} {\bf 98} (1935) 39.

\bibitem{Jungman:1995bz}
G.~Jungman, M.~Kamionkowski, A.~Kosowsky, and D.~N. Spergel, {\it {Cosmological
  parameter determination with microwave background maps}},  {\em Phys. Rev.}
  {\bf D54} (1996) 1332--1344,
  [\href{http://xxx.lanl.gov/abs/astro-ph/9512139}{{\tt astro-ph/9512139}}].

\bibitem{Tegmark:1996bz}
M.~Tegmark, A.~Taylor, and A.~Heavens, {\it {Karhunen-Loeve eigenvalue problems
  in cosmology: how should we tackle large data sets?}},  {\em Astrophys. J.}
  {\bf 480} (1997) 22, [\href{http://xxx.lanl.gov/abs/astro-ph/9603021}{{\tt
  astro-ph/9603021}}].

\bibitem{Piattella:2011fv}
O.~F. Piattella and D.~Bertacca, {\it {Gravitational potential evolution in
  Unified Dark Matter Scalar Field Cosmologies: an analytical approach}},  {\em
  Mod. Phys. Lett.} {\bf A26} (2011) 2277--2286,
  [\href{http://xxx.lanl.gov/abs/1103.0234}{{\tt arXiv:1103.0234}}].

\bibitem{Seo:2003pu}
H.-J. Seo and D.~J. Eisenstein, {\it {Probing dark energy with baryonic
  acoustic oscillations from future large galaxy redshift surveys}},  {\em
  Astrophys. J.} {\bf 598} (2003) 720--740,
  [\href{http://xxx.lanl.gov/abs/astro-ph/0307460}{{\tt astro-ph/0307460}}].

\bibitem{Wang:2006qt}
Y.~Wang, {\it {Dark energy constraints from baryon acoustic oscillations}},
  {\em Astrophys. J.} {\bf 647} (2006) 1--7,
  [\href{http://xxx.lanl.gov/abs/astro-ph/0601163}{{\tt astro-ph/0601163}}].

\bibitem{Wang:2007ht}
Y.~Wang, {\it {Differentiating dark energy and modified gravity with galaxy
  redshift surveys}},  {\em JCAP} {\bf 0805} (2008) 021,
  [\href{http://xxx.lanl.gov/abs/0710.3885}{{\tt arXiv:0710.3885}}].

\bibitem{Wang:2008zh}
Y.~Wang, {\it {Figure of Merit for Dark Energy Constraints from Current
  Observational Data}},  {\em Phys. Rev.} {\bf D77} (2008) 123525,
  [\href{http://xxx.lanl.gov/abs/0803.4295}{{\tt arXiv:0803.4295}}].

\bibitem{Wang:2010gq}
Y.~Wang, W.~Percival, A.~Cimatti, P.~Mukherjee, L.~Guzzo, {\em et.~al.}, {\it
  {Designing a space-based galaxy redshift survey to probe dark energy}},  {\em
  Mon.Not.Roy.Astron.Soc.} {\bf 409} (2010) 737,
  [\href{http://xxx.lanl.gov/abs/1006.3517}{{\tt arXiv:1006.3517}}].

\bibitem{Carbone:2010ik}
C.~Carbone, L.~Verde, Y.~Wang, and A.~Cimatti, {\it {Neutrino constraints from
  future nearly all-sky spectroscopic galaxy surveys}},  {\em JCAP} {\bf 1103}
  (2011) 030, [\href{http://xxx.lanl.gov/abs/1012.2868}{{\tt
  arXiv:1012.2868}}].

\bibitem{Carbone:2011bx}
C.~Carbone, A.~Mangilli, and L.~Verde, {\it {Isocurvature modes and Baryon
  Acoustic Oscillations II: gains from combining CMB and Large Scale
  Structure}},  {\em JCAP} {\bf 1109} (2011) 028,
  [\href{http://xxx.lanl.gov/abs/1107.1211}{{\tt arXiv:1107.1211}}].

\bibitem{Carbone:2011by}
C.~Carbone, C.~Fedeli, L.~Moscardini, and A.~Cimatti, {\it {Measuring the
  neutrino mass from future wide galaxy cluster catalogues}},
  \href{http://xxx.lanl.gov/abs/1112.4810}{{\tt arXiv:1112.4810}}.

\bibitem{DiPorto:2012ey}
C.~Di~Porto, L.~Amendola, and E.~Branchini, {\it {Simultaneous constraints on
  bias, normalization and growth index through power spectrum measurements}},
  \href{http://xxx.lanl.gov/abs/1201.2455}{{\tt arXiv:1201.2455}}.

\bibitem{Amendola:2011ie}
L.~Amendola, V.~Pettorino, C.~Quercellini, and A.~Vollmer, {\it {Testing
  coupled dark energy with next-generation large-scale observations}},
  \href{http://xxx.lanl.gov/abs/1111.1404}{{\tt arXiv:1111.1404}}.

\bibitem{DiPorto:2011jr}
C.~Di~Porto, L.~Amendola, and E.~Branchini, {\it {Growth factor and galaxy bias
  from future redshift surveys: a study on parametrizations}},  {\em Mon. Not.
  Roy. Astron. Soc.} {\bf 419} (2011) 985,
  [\href{http://xxx.lanl.gov/abs/1101.2453}{{\tt arXiv:1101.2453}}].

\bibitem{Kaiser:1987qv}
N.~Kaiser, {\it {Clustering in real space and in redshift space}},  {\em Mon.
  Not. Roy. Astron. Soc.} {\bf 227} (1987) 1--27.

\bibitem{Wang:2009gt}
Y.~Wang, {\it {Clarifying Forecasts of Dark Energy Constraints from Baryon
  Acoustic Oscillations}},  {\em Mod. Phys. Lett.} {\bf A25} (2010) 3093--3113,
  [\href{http://xxx.lanl.gov/abs/0904.2218}{{\tt arXiv:0904.2218}}].

\bibitem{Ballinger:1996cd}
W.~Ballinger, J.~Peacock, and A.~Heavens, {\it {Measuring the cosmological
  constant with redshift surveys}},  {\em Mon. Not. Roy. Astron. Soc.} {\bf
  282} (1996) 877--888, [\href{http://xxx.lanl.gov/abs/astro-ph/9605017}{{\tt
  astro-ph/9605017}}].

\bibitem{Marulli:2012na}
F.~Marulli, D.~Bianchi, E.~Branchini, L.~Guzzo, L.~Moscardini, {\em et.~al.},
  {\it {Cosmology with clustering anisotropies: disentangling dynamic and
  geometric distortions in galaxy redshift surveys}},
  \href{http://xxx.lanl.gov/abs/1203.1002}{{\tt arXiv:1203.1002}}.

\bibitem{Seljak:2000gq}
U.~Seljak, {\it {Analytic model for galaxy and dark matter clustering}},  {\em
  Mon. Not. Roy. Astron. Soc.} {\bf 318} (2000) 203,
  [\href{http://xxx.lanl.gov/abs/astro-ph/0001493}{{\tt astro-ph/0001493}}].

\bibitem{Marulli:2011jk}
F.~Marulli, M.~Baldi, and L.~Moscardini, {\it {Clustering and redshift-space
  distortions in interacting dark energy cosmologies}},  {\em Mon. Not. Roy.
  Astron. Soc.} (2011) [\href{http://xxx.lanl.gov/abs/1110.3045}{{\tt
  arXiv:1110.3045}}].

\bibitem{Marulli:2011he}
F.~Marulli, C.~Carbone, M.~Viel, L.~Moscardini, and A.~Cimatti, {\it {Effects
  of Massive Neutrinos on the Large-Scale Structure of the Universe}},  {\em
  Mon. Not. Roy. Astron. Soc.} {\bf 418} (2011) 346,
  [\href{http://xxx.lanl.gov/abs/1103.0278}{{\tt arXiv:1103.0278}}].

\bibitem{Tegmark:1997rp}
M.~Tegmark, {\it {Measuring cosmological parameters with galaxy surveys}},
  {\em Phys. Rev. Lett.} {\bf 79} (1997) 3806--3809,
  [\href{http://xxx.lanl.gov/abs/astro-ph/9706198}{{\tt astro-ph/9706198}}].

\bibitem{Kaiser:1996tp}
N.~Kaiser, {\it {Weak Lensing and Cosmology}},  {\em Astrophys. J.} {\bf 498}
  (1998) 26, [\href{http://xxx.lanl.gov/abs/astro-ph/9610120}{{\tt
  astro-ph/9610120}}].

\bibitem{Bartelmann:1999yn}
M.~Bartelmann and P.~Schneider, {\it {Weak Gravitational Lensing}},  {\em Phys.
  Rept.} {\bf 340} (2001) 291--472,
  [\href{http://xxx.lanl.gov/abs/astro-ph/9912508}{{\tt astro-ph/9912508}}].

\bibitem{Hu:1999ek}
W.~Hu, {\it {Power Spectrum Tomography with Weak Lensing}},  {\em Astrophys.
  J.} {\bf 522} (1999) L21--L24,
  [\href{http://xxx.lanl.gov/abs/astro-ph/9904153}{{\tt astro-ph/9904153}}].

\bibitem{Zaldarriaga:1996xe}
M.~Zaldarriaga and U.~Seljak, {\it {An all sky analysis of polarization in the
  microwave background}},  {\em Phys. Rev.} {\bf D55} (1997) 1830--1840,
  [\href{http://xxx.lanl.gov/abs/astro-ph/9609170}{{\tt astro-ph/9609170}}].

\bibitem{Zaldarriaga:1997ch}
M.~Zaldarriaga, D.~N. Spergel, and U.~Seljak, {\it {Microwave background
  constraints on cosmological parameters}},  {\em Astrophys. J.} {\bf 488}
  (1997) 1--13, [\href{http://xxx.lanl.gov/abs/astro-ph/9702157}{{\tt
  astro-ph/9702157}}].

\bibitem{Bond:1997wr}
J.~Bond, G.~Efstathiou, and M.~Tegmark, {\it {Forecasting cosmic parameter
  errors from microwave background anisotropy experiments}},  {\em Mon. Not.
  Roy. Astron. Soc.} {\bf 291} (1997) L33--L41,
  [\href{http://xxx.lanl.gov/abs/astro-ph/9702100}{{\tt astro-ph/9702100}}].

\bibitem{2010MNRAS.402.1330G}
J.~E. {Geach}, A.~{Cimatti}, W.~{Percival}, Y.~{Wang}, L.~{Guzzo},
  G.~{Zamorani}, P.~{Rosati}, L.~{Pozzetti}, A.~{Orsi}, C.~M. {Baugh}, C.~G.
  {Lacey}, B.~{Garilli}, P.~{Franzetti}, J.~R. {Walsh}, and M.~{K{\"u}mmel},
  {\it {Empirical H{$\alpha$} emitter count predictions for dark energy
  surveys}},  {\em Mon. Not. Roy. Astron. Soc.} {\bf 402} (Feb., 2010)
  1330--1338, [\href{http://xxx.lanl.gov/abs/0911.0686}{{\tt
  arXiv:0911.0686}}].

\bibitem{2010MNRAS.405.1006O}
A.~{Orsi}, C.~M. {Baugh}, C.~G. {Lacey}, A.~{Cimatti}, Y.~{Wang}, and
  G.~{Zamorani}, {\it {Probing dark energy with future redshift surveys: a
  comparison of emission line and broad-band selection in the near-infrared}},
  {\em Mon. Not. Roy. Astron. Soc.} {\bf 405} (June, 2010) 1006--1024,
  [\href{http://xxx.lanl.gov/abs/0911.0669}{{\tt arXiv:0911.0669}}].

\bibitem{1994MNRAS.270..245S}
I.~{Smail}, R.~S. {Ellis}, and M.~J. {Fitchett}, {\it {Gravitational Lensing of
  Distant Field Galaxies by Rich Clusters - Part One - Faint Galaxy Redshift
  Distributions}},  {\em Mon. Not. Roy. Astron. Soc.} {\bf 270} (Sept., 1994)
  245--+, [\href{http://xxx.lanl.gov/abs/astro-ph/}{{\tt astro-ph/}}].

\bibitem{Albrecht:2006um}
A.~J. Albrecht {\em et.~al.}, {\it {Report of the Dark Energy Task Force}},
  {\em arXiv:astro-ph/0609591} (2006)
  [\href{http://xxx.lanl.gov/abs/astro-ph/0609591}{{\tt astro-ph/0609591}}].

\bibitem{Eisenstein:1997ik}
D.~J. Eisenstein and W.~Hu, {\it {Baryonic Features in the Matter Transfer
  Function}},  {\em Astrophys. J.} {\bf 496} (1998) 605,
  [\href{http://xxx.lanl.gov/abs/astro-ph/9709112}{{\tt astro-ph/9709112}}].

\bibitem{Lewis:1999bs}
A.~Lewis, A.~Challinor, and A.~Lasenby, {\it {Efficient Computation of CMB
  anisotropies in closed FRW models}},  {\em Astrophys. J.} {\bf 538} (2000)
  473--476, [\href{http://xxx.lanl.gov/abs/astro-ph/9911177}{{\tt
  astro-ph/9911177}}].

\bibitem{Xu:2011bp}
L.~Xu, Y.~Wang, and H.~Noh, {\it {A Unified Dark Fluid with Constant Adiabatic
  Sound Speed and Cosmic Constraints}},
  \href{http://xxx.lanl.gov/abs/1112.3701}{{\tt arXiv:1112.3701}}.

\bibitem{Aviles:2011ak}
A.~Aviles and J.~L. Cervantes-Cota, {\it {The dark degeneracy and interacting
  cosmic components}},  {\em Phys. Rev.} {\bf D84} (2011) 083515,
  [\href{http://xxx.lanl.gov/abs/1108.2457}{{\tt arXiv:1108.2457}}].

\end{thebibliography}\endgroup

\end{document}